\documentclass{article}
\usepackage[english]{babel}
\usepackage[letterpaper,top=2cm,bottom=2cm,left=3cm,right=3cm,marginparwidth=1.75cm]{geometry}
\usepackage{caption}
\usepackage{subcaption}
\usepackage{amsmath}
\usepackage{graphicx}
\usepackage[colorlinks=true, allcolors=blue]{hyperref}
\usepackage{siunitx}
\usepackage{cancel}
\usepackage{graphicx}
\usepackage{pgfplots}
\usepackage{listings}
\usepackage{tikz}
\usepackage{hyperref}
\usepackage{float}
\usepackage{geometry}
\usepackage{lscape}
\usepackage{textcomp}
\usepackage{xcolor}
\usepackage{fancyhdr}
\usepackage{dirtytalk}
\usepackage[font=small]{caption}
\usepackage[numbers]{natbib}

\pgfplotsset{width=10cm,compat=1.9}
\usepgfplotslibrary{external}
\title{Revisiting Cont's Stylized Facts for Modern Stock Markets
\thanks{C.M.V.O., M.T.T.K., and B.F.T. were supported by MITRE’s Financial Innovation Lab. E.R.C. was supported by the MITRE PhD Fellowship in Computational Finance within the Complex Systems Center at the University of Vermont.\newline
©2023 The MITRE Corporation. ALL RIGHTS RESERVED. Approved for Public Release. Distribution unlimited, Case 23-3523.
}
}
\author{
    Ethan Ratliff-Crain\\
    \textit{Vermont Complex Systems Center} \\
    \textit{University of Vermont}\\
    Burlington, VT 05405, USA \\
    \textit{The MITRE Corporation}\\
    McLean, VA 22012, USA \\
    ethan.ratliff-crain@uvm.edu
    \and
    Colin M. Van Oort\\
    \textit{The MITRE Corporation}\\
    McLean, VA 22012, USA \\
    cvanoort@mitre.org
    \and
    James Bagrow\\
    \textit{Department of Mathematics and Statistics} \\
    \textit{University of Vermont}\\
    Burlington, VT 05405, USA \\
    james.bagrow@uvm.edu
    \and
    Matthew T. K. Koehler\\
    \textit{The MITRE Corporation}\\
    McLean, VA 22012, USA \\
    mkoehler@mitre.org
    \and
    Brian F. Tivnan\\
    \textit{The MITRE Corporation}\\
    McLean, VA 22012, USA \\
    \textit{Vermont Complex Systems Center} \\
    \textit{University of Vermont}\\
    Burlington, VT 05405, USA \\
    btivnan@mitre.org
}

\begin{document}

\maketitle
\begin{sloppypar}

\begin{abstract}
In 2001, Rama Cont introduced a now-widely used set of ‘stylized facts’ to synthesize empirical studies of financial price changes (returns), resulting in 11 statistical properties common to a large set of assets and markets. These properties are viewed as constraints a model should be able to reproduce in order to accurately represent returns in a market. It has not been established whether the characteristics Cont noted in 2001 still hold for modern markets following significant regulatory shifts and technological advances. It is also not clear whether a given time series of financial returns for an asset will express all 11 stylized facts. We test both of these propositions by attempting to replicate each of Cont's 11 stylized facts for intraday returns of the individual stocks in the Dow 30, using the same authoritative data as that used by the U.S. regulator from October 2018 -- March 2019. We find conclusive evidence for eight of Cont's original facts and no support for the remaining three.
Our study represents the first test of Cont's 11 stylized facts against a consistent set of stocks, therefore providing insight into how these stylized facts should be viewed in the context of modern stock markets.

\end{abstract}

\newpage
\lfoot{\tiny ©2023 The MITRE Corporation. ALL RIGHTS RESERVED. 
\newline Approved for Public Release. Distribution unlimited, Case 23-2969.}

\section{Introduction}
Financial markets feature numerous agents collectively establishing an evolving price for an asset. Despite the complex nature of this process over time and the specific details of the asset in question, scientists for decades have noted seeming commonalities that emerge across markets. These high-level characterizations are referred to as stylized facts, which are then used to inform and evaluate models of markets. One of the most, if not the most, widely attributed set of stylized facts of financial markets was written by Rama Cont and published in 2001 \cite{cont_empirical_2001}. Cont's review summarized research ranging over the decades prior, noting 11 seemingly common statistical properties of price changes (returns) across different markets and time frames. Cont argued these properties \say{should be viewed as constraints that a stochastic process has to verify in order to reproduce the statistical properties of returns accurately} \cite{cont_empirical_2001}(p. 233).

Markets in the 21$^{st}$ century have experienced a marked shift through increasing automation and responding regulatory changes \cite{angel_when_2014}\cite{budish_will_2019}\cite{johnson_abrupt_2013}\cite{kirilenko_flash_2017}\cite{tivnan_fragmentation_2020}\cite{van_oort_ecological_2022}. 
It has not been established whether the characteristics Cont \cite{cont_empirical_2001} noted in 2001 still hold for modern markets, nor whether the full set of stylized facts should be expected to hold for individual securities for a given time period.
In this study, we examine the most granular stock trading data publicly available over the time period of 18 Oct. 2018 -- 19 Mar. 2019, testing whether intraday stock returns in the modern U.S. stock market express each of Cont's 11 stylized facts.
As summarized Table \ref{table:fact_summary}, we find clear support for eight of the stylized facts and no support for the other three from the measures tested. Section \ref{literature} provides further motivation and background for the study. Section \ref{methods} details our data and general methodology, and Section \ref{results} gives our specific analyses and results for each of the stylized facts. Section \ref{discussion} discusses our findings and possible explanations for the three facts for which we did not find support. Finally, Section \ref{conclusion} summarizes takeaways from our results.

\section{Background}
\label{literature}
Cont's stylized facts \cite{cont_empirical_2001} drew from numerous past results of Cont and others over the second-half of the 1900's into the early 2000's. Some of the facts were demonstrated through results in the paper, others cited past results, and a couple (detailed in Section \ref{cont_facts}) do not appear to have been directly cited or reproduced in the paper. Cont \cite{cont_empirical_2001} argued for these properties to be used as constraints a model should be able to reproduce, and in the years since they have frequently used to benchmark the empirical relevance of agent-based models (ABMs) to real-world financial markets \cite{axtell_three_2005}. Cont himself was involved in these efforts \cite{ghoulmie_heterogeneity_2005}, and numerous ABMs have similarly replicated multiple stylized facts  \cite{bookstaber_agent-based_2018}\cite{johnson_abrupt_2013}\cite{mcgroarty_high_2019}\cite{preis_multi-agent-based_2006}\cite{tivnan_towards_2017}. More recently, Katahira et al. \cite{katahira_development_2019} gave a `speculation game' model with results reproducing 10 of the 11 facts (all except Fact \#3: Gain/Loss Asymmetry). In \cite{bookstaber_toward_2016} and \cite{tivnan_towards_2017}, the stylized facts were used to determine which parameters produce realistic return series. As Cont noted in 2001, \say{most currently existing models fail to reproduce all these statistical features at once} \cite{cont_empirical_2001} (p. 233). An implicit assumption is therefore that the stylized facts should each hold for a given return series.

Since Cont's set of stylized facts drew from a variety of results and research groups, no single asset (exchange rate, stock, index, etc.) was used for all 11 facts \cite{cont_empirical_2001}. The most facts tested by a single study that we have found is eight, done by Chakraborti et al. \cite{chakraborti_econophysics_2011} in their review on econophysics. In their summary of stylized facts, they provided example results using the French stock BNPP.PA from 2007 -- 2008, with evidence supporting Facts \#1, 2, 4, 5, 6, 7, 8, and 10. Determining the extent to which each of the stylized facts should be expected to hold for a given asset is important for understanding how the facts should be used in practice. If a given asset's returns over some time period cannot reliably be expected to exhibit all 11 facts, alternative interpretations should be considered. Perhaps stocks on average will exhibit all 11 properties, in which case comparing the expected value of a model's results to the expected value from empirical data could be more appropriate\footnote{This is the approach taken by Farmer et al. to validate their `Zero-Intelligence' model \cite{farmer_predictive_2005}.}.

How the stylized facts are exhibited under different constructions of time is also important to establish. Clock-time, meaning time as measured by a timestamp or date, is most widely represented in the stylized facts literature. As noted by Chakraborti et al. \cite{chakraborti_econophysics_2011}, this view inherently involves sampling, and the number of trades in a unit of time can vary widely period-to-period and stock-to-stock (Table \ref{table:num_trades}). Event-time --- using trades as the event in our case --- smooths this variability out, with one trade occurring per unit of time. An event-based view of time is also natural in the context of market simulations, making any differentiation between the stylized facts in clock-time versus event-time crucial to understand when benchmarking these models.


\subsection{Cont's Stylized Facts}
\label{cont_facts}
Below are Cont's stylized facts as given in \cite{cont_empirical_2001}:
\begin{enumerate}
    \item \textbf{Absence of autocorrelations:} \say{(Linear) autocorrelations of asset returns are often insignificant, except for very small intraday timescales ($\simeq 20$ minutes) for which microstructure effects come into play.}
    \item \textbf{Heavy tails:} \say{The (unconditional) distribution of returns seems to display a power-law or Pareto-like tail, with a tail index which is finite, higher than two and less than five for most data sets studied. In particular this excludes stable laws with infinite variance and the normal distribution. However the precise form of the tails is difficult to determine.}
    \item \textbf{Gain/loss asymmetry:} \say{One observes large drawdowns in stock prices and stock index values but not equally large upward movements.}
    \item \textbf{Aggregational Gaussianity:} \say{As one increases the timescale $\Delta{t}$ over which returns are calculated, their distribution looks more and more like the normal distribution. In particular, the shape of the distribution is not the same at different timescales.}
    \item \textbf{Intermittency:} \say{Returns display, at any time scale, a high degree of variability. This is quantified by the presence of irregular bursts in time series of a wide variety of volatility estimators.}
    \item \textbf{Volatility clustering:} \say{Different measures of volatility display a positive autocorrelation over several days, which quantifies the fact that high-volatility events tend to cluster in time.}
    \item \textbf{Conditional heavy tails:} \say{Even after correcting returns for volatility clustering (e.g. via GARCH-type models), the residual time series still exhibit heavy tails. However, the tails are less heavy than in the unconditional distribution of returns.}
    \item \textbf{Slow decay of autocorrelation in absolute returns:} \say{The autocorrelation function of absolute returns decays slowly as a function of the time lag, roughly as a power law with an exponent $\beta\in[0.2,0.4]$. This is sometimes interpreted as a sign of long-range dependence.}
    \item \textbf{Leverage effect:} \say{Most measures of volatility of an asset are negatively correlated with the returns of that asset.}
    \item \textbf{Volume/volatility correlation:} \say{Trading volume is correlated with all measures of volatility.}
    \item \textbf{Asymmetry in timescales:} \say{Coarse-grained measures of volatility predict fine-scale volatility better than the other way around.}
\end{enumerate}

As framed by Cont's facts, financial returns overall are heavy-tailed, not independent and identically distributed (iid), and characterized by correlations and clustering in behavior. Price changes themselves are not predicted by any of the stylized facts. Magnitudes of changes, seen as measures of volatility, are found to have nontrivial correlations and relationships with previous behavior. The lack of clear persisting signal on the raw returns is detailed in Fact \#1, measured as a lack of linear autocorrelation in returns. This fact is reproduced in the stylized facts paper \cite{cont_empirical_2001} for event-time returns of the stock KLM and for the USD/Yen exchange rate. Nonzero autocorrelation function (ACF) values at the first lag are found in these and many other intraday results in general, with possible explanations proposed such as the `bid-ask bounce', nonsynchronous trading effects, and partial price adjustment \cite{anderson_autocorrelation_2013}. The effect is found to decay to roughly zero within 15-minutes by Cont and others in \cite{arneodo__1998}\cite{cont_scaling_1997}\cite{todorova_power_2011}.

Some nonlinear transformations of returns, such as taking their absolute or squared values, provide measures of the magnitude of price changes. These volatility measures are found to exhibit persistent positive autocorrelation, in contrast to the linear ACF just discussed. Cont et al. \cite{cont_scaling_1997}\cite{cont_scaling_1997-1} found the absolute 5-minute returns of S\&P 500 futures had ACF values starting above 0.1 and not going below zero for at least 100 lags. Similar results were found in \cite{chakraborti_econophysics_2011}\cite{cont_volatility_2007}\cite{liu_correlations_1997}\cite{liu_statistical_1999}\cite{mikosch_long-range_2002}\cite{muller_statistical_1990}. Explicit power-law fits are given for the decay of autocorrelation in squared and absolute returns in \cite{cont_scaling_1997-1}\cite{liu_correlations_1997}\cite{liu_statistical_1999}. Power-law decay of absolute autocorrelation implies volatility exhibits long memory, or is `long-range correlated', and we would in that case expect the correlation to asymptotically go to zero as the lags increase \cite{mantegna_introduction_1999}.

The variability of returns is well documented and leads to the second stylized fact: return distributions' heavy tails. Trying to determine the precise distributional form of returns and their tails is a `favorite pastime' (as Cont put it) in the literature \cite{cont_empirical_2001}\cite{gopikrishnan_inverse_1998}\cite{gopikrishnan_scaling_2000}\cite{malevergne_empirical_2005}. Consistently agreed upon, however, is that returns exhibit kurtosis greater than that of a normal distribution \cite{andersen_intraday_1997}\cite{andersen_distribution_2001}\cite{bollerslev_arch_1992}\cite{chakraborti_econophysics_2011}\cite{cont_empirical_2001}\cite{cont_scaling_1997}\cite{cont_scaling_1997-1}\cite{muller_statistical_1990}\cite{plerou_economic_2000}\cite{potters_financial_1998}. Excess kurtosis indicates a distribution has heavier tails and a higher peak than a normal distribution \cite{decarlo_meaning_1997}. This measure has been frequently used by the financial ABM literature since 2001 to argue for the empirical relevance of market models \cite{bookstaber_toward_2016}\cite{ghoulmie_heterogeneity_2005}\cite{katahira_development_2019}\cite{mcgroarty_high_2019}\cite{tivnan_towards_2017}\cite{tivnan_towards_2021}.

In empirical results, returns were found to show excess kurtosis for timescales up to multiple days, but kurtosis was found to decrease overall with timescale \cite{andersen_intraday_1997}\cite{chakraborti_econophysics_2011}\cite{cont_scaling_1997}\cite{cont_scaling_1997-1}\cite{potters_financial_1998}\cite{muller_statistical_1990}. This property of `aggregational Gaussianity' (Fact \#4) was shown by Chakraborti et al. \cite{chakraborti_econophysics_2011} to occur more quickly in event-time than in clock-time, explained as event-time correcting for some volatility versus clock-time returns. Tail-heaviness decreasing but not necessarily disappearing through methods of volatility correction is summarized in Fact \#7, `conditional heavy tails'. Bollerslev et al. \cite{bollerslev_arch_1992} detail the `kurtosis problem' of fat-tails remaining in the residuals after applying autoregressive conditional heteroskedasticity (ARCH)-type models to stock returns. In \cite{andersen_intraday_1997}, Andersen and Bollerslev normalized 5-minute DEM/USD FX returns by an estimate of the average daily volatility pattern and reduced the kurtosis from 21.5 to 15.8. In Andersen et al. \cite{andersen_distribution_2001}, the realized standard deviation of 5-minute returns were used to normalize daily returns of Dow 30 stocks, resulting in nearly normal kurtosis for the normalized distributions.

Taken together, returns' heavy tails and volatility clustering lead to the characteristic that returns irregularly display periods of high volatility separated by long periods of relative calm. This `intermittency' is Fact \#5, and in much of the literature it is discussed as being visibly apparent in the returns series \cite{cont_empirical_2001}\cite{katahira_development_2019} or following directly from these other facts \cite{cont_empirical_2001}. Cont \cite{cont_empirical_2001} discusses the multifractal model as a possible explanation of intermittency, suggesting possible multiplicative processes operating across multiple timescales. Arneodo et al. \cite{arneodo__1998} provide evidence of this, arguing for a multiplicative cascade of information from coarse timescales to finer timescales. M\"uller et al. \cite{muller_volatilities_1997} and Zumbach and Lynch \cite{zumbach_heterogeneous_2001} presented evidence of this `asymmetry in timescales' effect\footnote{Asymmetry in timescales (Fact \#11) has also been referred to as the Zumbach effect \cite{el_euch_zumbach_2020} for the findings in Zumbach and Lynch \cite{zumbach_heterogeneous_2001}.} (Fact \#11) by measuring the correlation between fine volatility and lagged coarse volatility. These studies used different measures of volatility, but similarly found the coarse volatility predicted fine volatility better than the other way around. In similar analysis, Gen\c{c}ay et al. \cite{gencay_information_2004} found low volatility at a long timescale was likely followed by low volatility at a shorter timescale whereas high volatility did not necessarily show this same `vertical dependence'. 

Fact \#9, the `leverage effect', expects a negative relationship between volatility and returns. Citing results from Bouchaud et al. \cite{bouchaud_leverage_2001} and Pagan \cite{pagan_econometrics_1996}, Cont specifically describes this effect as showing a negative correlation between returns and subsequent volatility, suggesting negative returns lead to increased volatility. Correlation of volatility with subsequent returns was negligible, meanwhile. Taken with Fact \#11, the leverage effect and asymmetry in timescales have motivated the search for statistical models  \cite{blanc_quadratic_2017} that can capture `time-reversal asymmetry' (TRA): financial time series do not behave the same forwards as they would if played in reverse \cite{zumbach_heterogeneous_2001}. Chicheportiche and Bouchaud \cite{chicheportiche_fine-structure_2014} provided empirical evidence of asymmetry in timescales and the leverage effect using the daily prices of 280 stocks from 2000 -- 2009. They showed general quadratic autoregressive (QARCH) models were able to capture TRA but overestimated the effect. Blanc et al. \cite{blanc_quadratic_2017} introduced Quadratic Hawkes models that displayed TRA, calibrating their model on 5-minute returns of 133 stocks from 2000 -- 2009. They found evidence for the Zumbach effect in the empirical data but did not find a significant leverage effect at this timescale. Variations of the leverage effect have also been noted elsewhere in the literature, and negative (or even nonzero) correlation between returns and observed volatility is not always found \cite{ait-sahalia_leverage_2013}.

According to Fact \#10, volatility is positively correlated with trading volume. Clark \cite{clark_subordinated_1973} noted this relationship as far back as the 1970's when examining cotton prices. The relationship has been measured over the years through various means, including taking the correlation between shares traded and absolute returns over a period of time \cite{jain_dependence_1988} and measuring return variance as a function of the number of trades \cite{chakraborti_econophysics_2011}\cite{lamoureux_heteroskedasticity_1990}\cite{plerou_economic_2000}\cite{silva_stochastic_2007}. It has also been proposed that long-range autocorrelation of trading volume leads to volatility clustering \cite{plerou_economic_2000}.

Finally, `gain/loss asymmetry', the 3$^{rd}$ stylized fact, is perhaps the least clear to interpret from the detail given by Cont \cite{cont_empirical_2001}. This was summarized as larger drawdowns being seen than upward movements for stock prices and index values. It is possible this is referring to results given in \cite{cont_empirical_2001} showing negative skewness in the return distributions for S\&P 500 futures, Dollar/DM Futures, and Dollar/Swiss Franc futures, each at 5-minute timescales. The skew of a distribution implies something slightly different from the fact as summarized by Cont, however, as it does not necessarily tell you anything about which tail has larger values. Other studies have also found positive skew rather than negative \cite{shakeel_stylized_2021}. Some of the literature since Cont \cite{cont_empirical_2001} has examined `gain/loss asymmetry' from another direction, looking at the amount of time it takes to see a gain versus a loss above a certain magnitude. This `inverse statistic' has been used to show that a stock index will typically achieve a loss more quickly than a gain of the same magnitude \cite{jensen_inverse_2003}, but the same property was not found for individual stocks \cite{johansen_optimal_2006}, with correlated downward movements across stocks proposed as an explanation for why the phenomenon could arise in indices \cite{siven_multiscale_2009}.

\subsection{The National Market System}
\label{nms}

It is possible that some of the stylized facts no longer hold much descriptive power due to structural changes to modern markets. The U.S. stock market, known as the National Market System (NMS), has had numerous regulatory and technological changes this century. Trades in the NMS occur through the matching of buyers and sellers of a stock at a given price point. This can occur on stock exchanges or off-exchange through brokers, peer-to-peer trading, or at Alternative Trading Systems (ATSs). Each step in this process is increasingly automated, from order submission by the trader to matching and execution on the trading venue. In response to this development, the Securities and Exchange Commission (SEC) in 2005 adopted Regulation NMS, a new set of rules aimed at modernizing the U.S. stock market and providing increased protections for automated orders. As a result, the market has become more connected and more fragmented, with much higher trading volumes overall but a smaller portion of trades being executed by the stock exchanges than at the turn of the century.

\begin{figure}[t]
\centering
\includegraphics[width=1\linewidth]{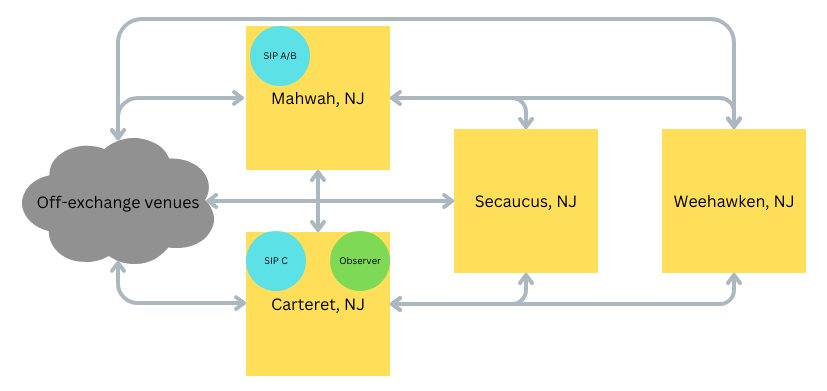}
\caption{The U.S. National Market System (NMS) circa 2018. The data collected here was collected from the SIP feeds by Thesys, the `observer' in Carteret. Image recreated from Van Oort et al. \cite{van_oort_ecological_2022}}
\label{fig:nms}
\end{figure}

As-of 2018, there were 13 stock exchanges split across four geographic locations in northern New Jersey, and three more exchanges have been added by the time of this writing. If we include ATSs and OTC firms, the number of venues totals in the hundreds \cite{finra_otc_nodate}\cite{finra_trading_nodate}. Per data reporting requirements from Regulation NMS, messages and trades from this array of venues are consolidated by the three Security Information Processor (SIP) `tapes'. Exchanges and off-exchange venues must report trades to the SIP tape corresponding to the traded security based on its listing exchange. This is diagrammed at a high level in Fig. \ref{fig:nms}. Roughly 30\% of trades in our data occurred off-exchange, a portion that has increased to roughly 45\% today \cite{cboe_global_markets_cboe_nodate}. Tivnan et al. \cite{tivnan_fragmentation_2020} give a more detailed summary of the market's infrastructure circa 2016, with evidence of impact from market fragmentation on the prices acted on by traders.

Each type of trading venue has its own specific rules and data-reporting requirements, and each venue reports trades with some (small but nonnegligible) allowed latency\footnote{Bartlett and McCrary \cite{bartlett_iii_how_2019} found trades of Dow 30 stocks were processed by the SIPs 24ms on average after they were recorded by exchange matching-engines over the period of 6 Aug. 2015 -- 30 Jun. 2016. Off-exchange trades, meanwhile, are allowed up to 10 seconds to report to Trade Reporting Facilities (TRFs) which then report the trades to the SIPs \cite{finra_trade_nodate}.}. Systems executing automated orders are required to sync their clocks to within 50ms of the time maintained by the National Institute of Standards and Technology (NIST) in the US \cite{lombardi_synchronizing_2020}. This may seem a short amount of time, except the NMS is increasingly operating at speeds approaching the speed of light, with reaction times measured in microseconds or even nanoseconds \cite{angel_when_2014}. Driven by algorithmic trading, a stock price can have an extreme rise or drop and rebound back to nearly its original level in less than a blink of the eye \cite{johnson_abrupt_2013}. Andersen et al. \cite{andersen_distribution_2001} noted the median time between trades was 23.1 seconds on average for Dow 30 stocks in 1993-1998. For contrast, in our sample, the median inter-trade duration for Dow 30 stocks was 693.77$\mu$s on average\footnote{Inter-trade durations are broken down in Table \ref{table:interarrival_stats} in our supplementary material}, a reduction of more than $10^4$ in magnitude. Given these modern developments, it is vital to understand whether the high-level characteristics of the market observed by Cont and others circa 2001 still hold in the market we have today.

\section{Methodology}
\label{methods}
We test each of Cont's 11 stylized facts on U.S. stock data from 18 Oct. 2018 -- 19 Mar. 2019 (103 trading days). We specifically looked at the individual stocks included in the Dow Jones Industrial Average as-of October 2018\footnote{The list of symbols is: AAPL, AXP, BA, CAT, CSCO, CVX, DIS, DWDP, GS, HD, IBM, INTC, JNJ, JPM, KO, MCD, MMM, MRK, MSFT, NKE, PFE, PG, TRV, UNH, UTX, V, VZ, WBA, WMT, XOM.} \cite{doorn_heres_2018}. 
The Dow Jones Industrial Average, commonly referred to as the Dow 30, is a well-known index dating back more than 100 years. The index consists of an evolving set of 30 stocks claimed to represent all industries in the U.S. stock market except for transportation and utilities \cite{indices_dow_nodate}. The Dow 30 is widely used as an indicator of the economy's performance more broadly, and the index and its underlying stocks are frequently used in studies of the stock market \cite{andersen_distribution_2001}\cite{bartlett_iii_how_2019}\cite{jensen_inverse_2003}\cite{malevergne_empirical_2005}\cite{tivnan_fragmentation_2020}. Thus, while this group of stocks is somewhat arbitrary, it gives an established cross-section of highly-traded stocks in the NMS.

Our data contains all trades reported for these symbols over the date range. The data was provided by Thesys Group Inc., which acted as a tape consolidator in the NMS and was the sole data provider for the SEC's MIDAS in this time period \cite{tivnan_fragmentation_2020}. The Thesys data was collected in the Nasdaq data center in Carteret, NJ (Fig. \ref{fig:nms}). Due to the latency and limitations on clock synchronization mentioned in Section \ref{nms}, the exact order of trades in the data is not definitive\footnote{And, in fact, a definitive ordering is not possible due to special relativity \cite{angel_when_2014}\cite{einstein_electrodynamics_1905}\cite{tivnan_fragmentation_2020}.}, but our time series provides the perspective of an observer located in Carteret, NJ, viewing the events of the market as reported by the SIPs.

We consider two different views of time when constructing our return time series: clock-time based on timestamp and event time, with trades as the event. In either view, in order to aggregate to any level that is more granular than a single trade per unit of time, we define $P\left(t,\Delta{t}\right)$ as the price of the last trade to occur in that time period. In clock-time, time points with no trades will be assumed to have the same price as the previous time point that had at least one trade, as no new price information has been received since then. Let $X\left(t, \Delta{t}\right)=\log{P\left(t,\Delta{t}\right)}$, the log-price. The log-return at time $t$ and timescale $\Delta{t}$ is defined as $r\left(t,\Delta{t}\right)=X\left(t,\Delta{t}\right)-X\left(t-1,\Delta{t}\right)$. Any reference to returns going forward should be interpreted as meaning log-returns. Note that time points with no trades will have a return of 0. We measure volatility as the absolute returns $|r\left(t,\Delta{t}\right)|$.

We limit our time series to the trading day (9:30am - 4:00pm ET), filtering out ’after-hours’ trading activity. We also filter out the batch auctions that start and end each day. In our price time series then, the last price before the close from one trading day will be followed immediately by the first price after the open of the next trading day. Due to this construction, an overnight return between $t_1=$ 16:00 on a given day and $t_2=$ 9:30 the following day is characteristically different from a return between two sequential prices within the same trading day. Therefore, we only consider returns $r(t,\Delta{t})$ such that $t$ and $t-\Delta{t}$ are within the range 9:30 -- 16:00 of the same trading day. If the last period of any day is incomplete (e.g. $\Delta{t}=50Min$ would result in a 40-minute period at the end of the day), we remove that return from our series. It is worth noting that the number of data points in a day will vary inversely with the timescale. Most stocks will have tens if not hundreds of thousands of trades per day, whereas there are 390 minutes in a 6.5-hour trading day\footnote{Financial markets notably have a daily pattern of trading and volatility characterized by high activity towards the ends of the day and relatively low activity in the middle of the day \cite{wood_investigation_1985}\cite{andersen_intraday_1997}\cite{malevergne_empirical_2005}\cite{dacorogna_geographical_1993}. We considered a normalization method to correct for this intraday seasonality in case it affected or obscured our analysis on the stylized facts. Our main results did not change after this normalization, however, and so we focus on the raw (un-normalized) returns for all facts except Fact \#7, which explicitly examines the effects of normalization on the heavy tails of return distributions. Our other results on the normalized returns are given in our supplementary material (Section \ref{normailized}).}.

Consistent with much of the results and references given by Cont \cite{cont_empirical_2001}, the main tools used in our analysis are correlation, the calculation of moments, and describing the distributions of events. We utilize Pearson sample correlation, denoted as $\text{corr}(x, y)$ going forward, to calculate the correlation values. In order to judge what is a consistent, nontrivial feature of the returns in our sample, we look at the extent a result shows a consistent signal across symbols, with that signal differing from what is observed for randomly generated `white noise' returns. To generate an instance of white noise returns, we create 103 days of prices whose trade-level returns are iid, normally distributed ($N\left(0,0.0001\right)$), with 250,000 trades randomly (uniformly) distributed throughout each day. By generating 100 of these 103-day price series and calculating results for each, we get a threshold of what type of behavior arises from random white noise.

\section{Results}
\label{results}
\begin{table}[t]
\centering
\begin{tabular}{|l|l|l|l|}
\hline
Fact \# & Fact Name &  Clock-time &  Event-time \\
\hline
1 & Lack of linear ACF & X & X \\
2 & Heavy tails & X & X \\
3 & Gain/Loss asymmetry & & \\
4 & Aggregational Gaussianity & X & X \\
5 & Intermittency & X & X \\
6 & Volatility Clustering & X & X \\
7 & Conditional heavy tails & X & X \\
8 & Slow decay of abs. ACF & X & X \\
9 & Leverage effect & & \\
10 & Volume/volatility corr. & X &  \\
11 & Asymmetry in timescales &  & \\
\hline
\end{tabular}
\caption{Breakdown of which facts we found evidence for in clock-time and event-time. `X' marks indicate strong evidence found for a fact in the respective timescale.}
\label{table:fact_summary}
\end{table}

Before getting into the specific results for each fact, Table \ref{table:fact_summary} gives a high-level summary of our findings. We find clear evidence for eight of the 11 stylized facts. Nuances to these results are unpacked in detail in the below subsections.


\subsection{Linear Autocorrelation of Returns}
\label{acf}
\begin{figure}[H]
\centering
\begin{subfigure}[t]{.48\textwidth}
  \centering
  \includegraphics[width=1\linewidth]{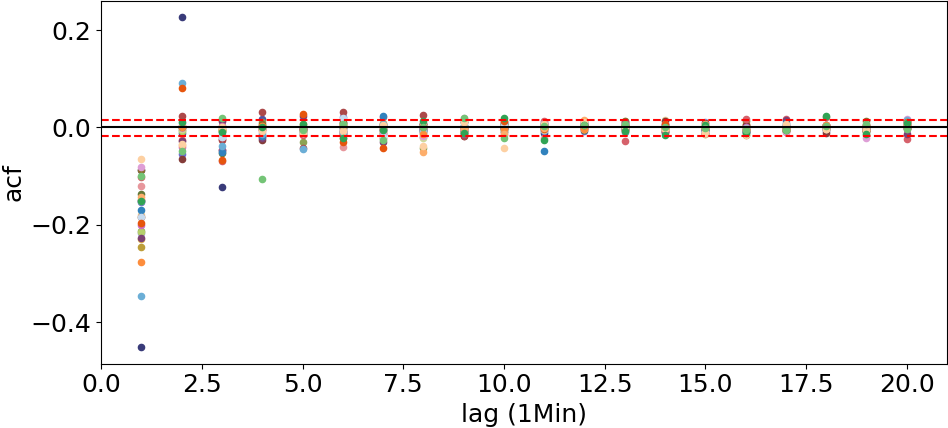}
\end{subfigure}
\hspace{1em}
\begin{subfigure}[t]{.48\textwidth}
  \centering
  \includegraphics[width=1\linewidth]{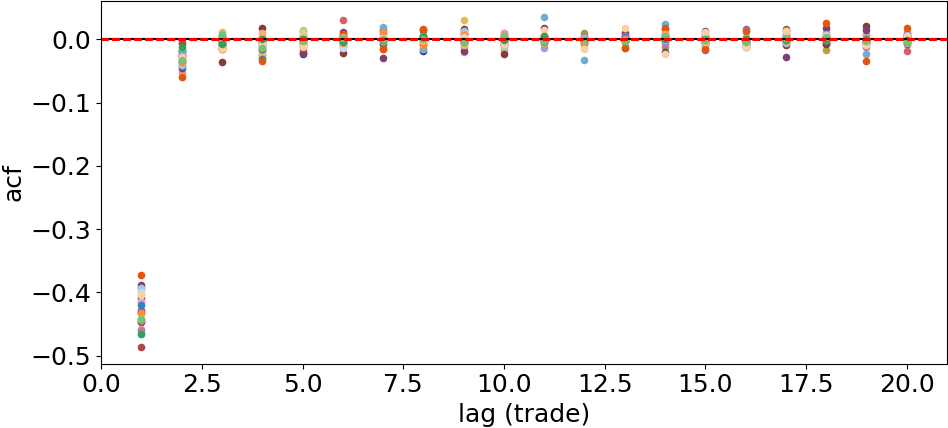}
\end{subfigure}
\begin{subfigure}[t]{\textwidth}
  \centering
  \includegraphics[width=1\linewidth]{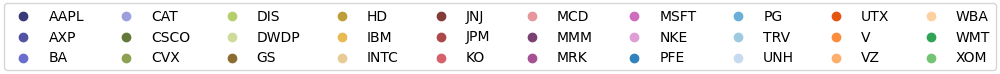}
\end{subfigure}
\caption{Linear autocorrelation of returns. Please note, the color legend shown here is used for all result plots.}
\label{fig:autocorrelation_d}
\end{figure}

Stylized Fact \#1 expects that linear autocorrelations of returns are \say{insignificant, except for very small intraday timescales ($\simeq 20$ minutes) for which microstructure effects come into play.} \cite{cont_empirical_2001}. The linear autocorrelation function (ACF) is 
$$C(\tau, \Delta{t})=\text{corr}(r(t,\Delta{t}), r(t+\tau,\Delta{t})).$$
Shown in Fig. \ref{fig:autocorrelation_d}, linear autocorrelation of the 1-minute returns in our sample is rather weak and difficult to differentiate from white noise past lags of about eight minutes. The first-lag ACF values are negative and outside the range of the white noise returns. Past the first lag, the sign of the ACF varies by symbol. The magnitudes of the correlations go to zero, although some symbols dip outside the range of white noise at later lags.

At the trade-level, we see negative ACF values in the first lag, characteristic of the so-called `bid-ask bounce' \cite{anderson_autocorrelation_2013}\cite{cont_empirical_2001}. Starting between -0.35 and -0.5 at the first lag, the ACF goes to zero within the next few lags. By lag $\tau=3$, at least some of the symbols have positive ACF while others have negative values. Due to the large number of observations in this timescale, the white noise levels are very small, and the observed values fall outside those thresholds. All values beyond the first two lags are below 0.05 in magnitude, and most are below 0.02. We rely on this and the fact the sign of the ACF varies by the symbol and lag to argue that linear dependence is unpredictable and weak past the first couple lags.

\subsection{Heavy Tails and Aggregational Gaussianity}
\label{heavy_tails}
Cont's 2$^{nd}$ fact expects return distributions to exhibit heavy tails. As done by Cont and collaborators in \cite{cont_empirical_2001}\cite{cont_scaling_1997}\cite{cont_scaling_1997-1}\cite{potters_financial_1998}, we examine the fourth central moment of returns, kurtosis. The kurtosis provides a measure of the tailedness of a distribution. We calculate kurtosis as defined below:
$$K(\Delta{t})=\frac{\langle(r(t,\Delta{t})-\langle r(t,\Delta{t})\rangle)^4\rangle}{\sigma(\Delta{t})^4}-3,$$
where $\sigma(\Delta{t})^2$ is the variance of the returns. Note that this definition of kurtosis subtracts 3 in order for the normal distribution to have a kurtosis of zero. Positive kurtosis means a distribution displays a sharper peak and heavier tails than a normal distribution. Similar to Cont et al. \cite{cont_scaling_1997-1}\cite{potters_financial_1998}, we plot the kurtosis as a function of $j\Delta{t}$, expecting positive $K(j\Delta{t})$ that decreases as $j$ increases.

\begin{figure}[H]
\centering
\begin{subfigure}[t]{.48\textwidth}
    \centering
    \includegraphics[width=1\linewidth]{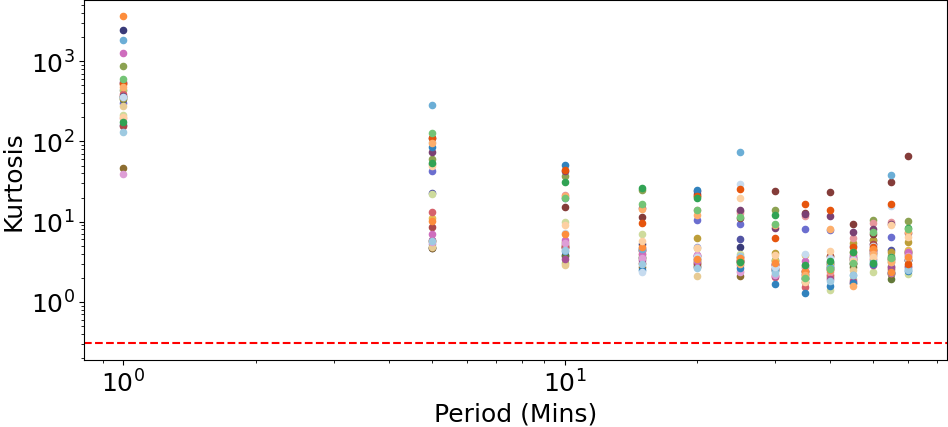}
\end{subfigure}%
\hspace{1em}
\begin{subfigure}[t]{.48\textwidth}
    \centering
    \includegraphics[width=1\linewidth]{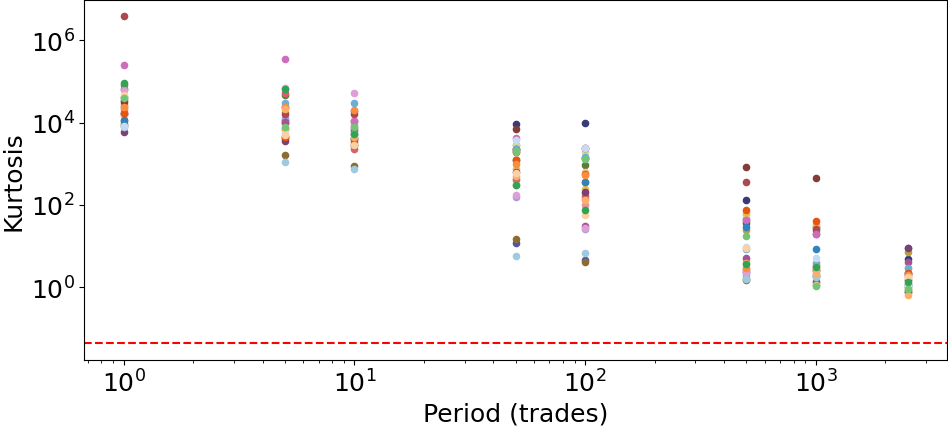}
\end{subfigure}%
\caption{Kurtosis as a function of timescale.}
\label{fig:kurtosis}
\end{figure}

We see the expected excess kurtosis in clock-time and event-time, as shown in Fig. \ref{fig:kurtosis}. For $\Delta{t}=1Min$, the kurtosis values range in magnitude from $10$ to more than $10^3$ depending on the symbol, with an overall negative trend as the timescale increases. All symbols have $K(\Delta{t}=1Min)>K(\Delta{t}=60Min)$, although kurtosis stays above the range of the Gaussian white noise returns, as shown by the red line in the plots. Both these features are as expected by Fact \#4, `aggregational Gaussianity', with past results finding return kurtosis to decrease with timescale but stay positive for timescales of up to multiple days \cite{cont_scaling_1997-1}\cite{muller_statistical_1990}\cite{potters_financial_1998}.

The event-time returns exhibit heavy tails and aggregational Gaussianity even more clearly than clock-time. The kurtosis values of the trade-level returns range from $10^3$ to more than $10^6$ depending on the symbol. Through aggregation in event-time, we see kurtosis decrease in a nearly monotonic fashion, going below $K(N)=10$ for all symbols for $N=2,500$. In clock-time, there are this many trades in less than a half-hour on average for even our lowest-traded stock (TRV, Table \ref{table:num_trades}). The event-time returns therefore show a more consistent decrease as a function of timescale, although returns exhibit timescales for the largest timescales tested in both clocks ($\Delta{t}=60Min$ in clock-time, 2,500 trades in event-time). Volume provides a proxy for volatility (as discussed later for Fact \#10 in Section \ref{volume_vol}), and as such viewing returns in event-time is one method of correcting for volatility in the return series. Through this method of volatility correction, we see the return distributions converge more quickly towards the normal distribution as a function of timescale.

\begin{table}
\centering
\begin{tabular}{|l|r|r|r|r|r|r|}
\hline
symbol & mean & total & max & min\\
\hline
AAPL & 252,807.20 & 26039142 & 721853 & 117999 \\
AXP & 33,493.62 & 3449843 & 73697 & 8966 \\
BA & 62,289.28 & 6415796 & 413938 & 21221 \\
CAT & 50,528.90 & 5204477 & 182012 & 24837 \\
CSCO & 120,418.70 & 12403126 & 263445 & 47746 \\
CVX & 52,099.22 & 5366220 & 94161 & 32564 \\
DIS & 62,727.27 & 6460909 & 193100 & 24834 \\
DWDP & 71,193.21 & 7332901 & 167513 & 26761 \\
GS & 40,096.39 & 4129928 & 118901 & 16930 \\
HD & 51,516.81 & 5306231 & 126587 & 27976 \\
IBM & 48,087.12 & 4952973 & 160827 & 20443 \\
INTC & 138,532.57 & 14268855 & 393166 & 56454 \\
JNJ & 65,590.05 & 6755775 & 412483 & 26550 \\
JPM & 111,131.54 & 11446549 & 253560 & 51990 \\
KO & 74,571.21 & 7680835 & 242201 & 29394 \\
MCD & 36,356.89 & 3744760 & 102455 & 17476 \\
MMM & 27,248.96 & 2806643 & 78667 & 11862 \\
MRK & 74,048.91 & 7627038 & 129195 & 26488 \\
MSFT & 235,426.20 & 24248899 & 539691 & 94358 \\
NKE & 48,301.92 & 4975098 & 154043 & 20798 \\
PFE & 105,104.29 & 10825742 & 200288 & 44942 \\
PG & 69,254.25 & 7133188 & 164793 & 28503 \\
TRV & 16,988.17 & 1749781 & 29834 & 4559 \\
UNH & 42,717.37 & 4399889 & 101063 & 13879 \\
UTX & 42,749.07 & 4403154 & 129023 & 21438 \\
V & 76,354.45 & 7864508 & 150305 & 35296 \\
VZ & 86,879.92 & 8948632 & 186526 & 42620 \\
WBA & 45,936.83 & 4731494 & 105880 & 21109 \\
WMT & 62,551.35 & 6442789 & 127323 & 35367 \\
XOM & 83,390.65 & 8589237 & 166697 & 51635 \\
\hline
all & 693.77 & 7,856,813.73 & 206,107.57 & 33,499.83 \\
\hline
\end{tabular}
\caption{Stats on the number of trades per day for Dow 30 stocks in our sample.}
\label{table:num_trades}
\end{table}

\begin{figure}[H]
\centering
\begin{subfigure}[t]{.48\textwidth}
    \centering
    \includegraphics[width=1\linewidth]{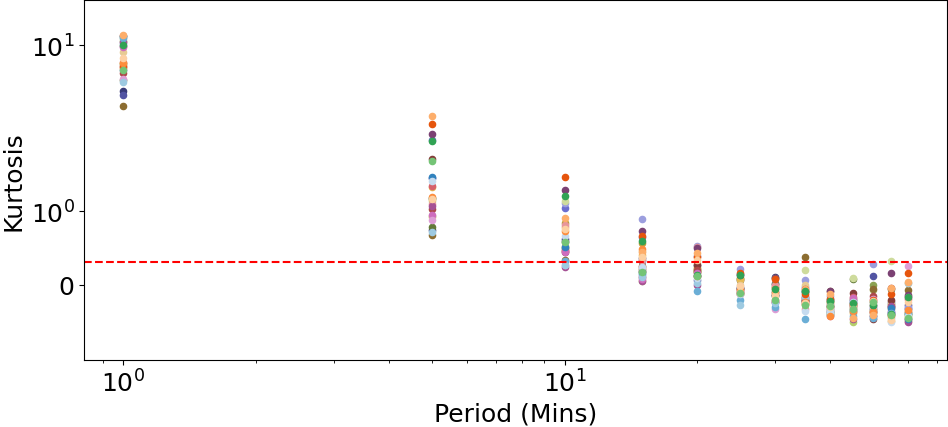}
\end{subfigure}%
\hspace{1em}
\begin{subfigure}[t]{.48\textwidth}
    \centering
    \includegraphics[width=1\linewidth]{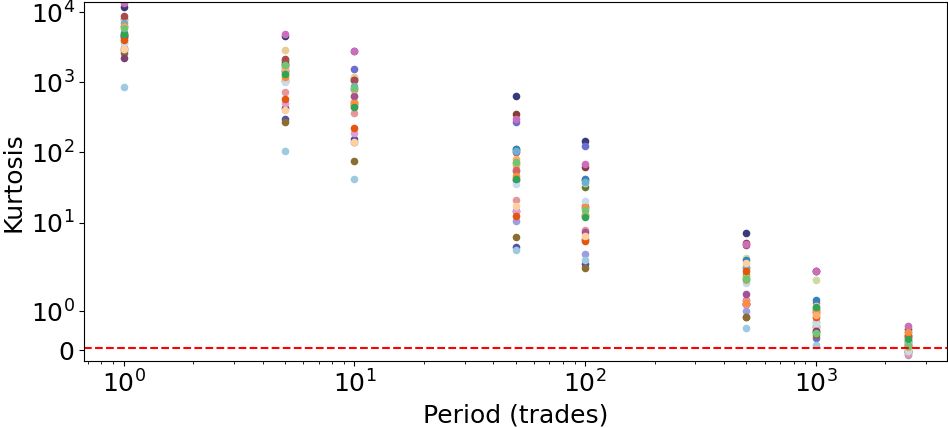}
\end{subfigure}%
\caption{Kurtosis of returns normalized by the daily standard deviation and daily volatility pattern.}
\label{fig:kurtosis_norm}
\end{figure}

This latter finding is in keeping with Cont's 7$^{th}$ stylized fact, conditional heavy tails. We can further test this property by normalizing the returns by the stock's daily volatility and its average volatility at a time of day. Specifically, we first normalize the returns from each day $T$ by the standard deviation of returns on that day at the given timescale, $\sigma\left(T, \Delta{t}\right)$. Say that $r_T\left(t, \Delta{t}\right)$ indicates a return is for a time-point on day $T$. Then, let
$$r'_T\left(t,\Delta{t}\right)=\frac{r_T\left(t,\Delta{t}\right)}{\sigma\left(T, \Delta{t}\right)}$$
Next, for a clock-timescale $\Delta{t}$, let
$$v\left(t,\Delta{t}\right)=\langle r'_T\left(t,\Delta{t}\right)\rangle_{T=1,2,...103},$$
the average absolute return at time $t$ across the 103-day sample for the stock, and let 
$$\hat{r}_T\left(t,\Delta{t}\right)=\frac{r'_T\left(t,\Delta{t}\right)}{v\left(t,\Delta{t}\right)}.$$ 
We do not do this latter normalization in event-time, since event-time is already correcting for different activity levels at different times of day by fixing the number of trades in a bucket. In other words, for an event-timescale $N$, $\hat{r}$ is simply:
$$\hat{r}_T\left(t,N\right)=r'_T\left(t,N\right).$$
This gives us two different views of normalized returns in addition to the raw returns calculated in either timescale.
The kurtosis for the normalized returns as a function of timescale is shown in Fig. \ref{fig:kurtosis_norm}. We see that our normalization of the returns does reduce the kurtosis from the unconditional returns while still leaving some excess kurtosis at small timescales. This is exactly as expected from Cont's description of conditional heavy tails. As we increase the timescales, both the clock- and event-time kurtosis values go to zero and even slightly negative. The average kurtosis of the normalized returns is below one for $\Delta{t}\geq15Min$ and within the range of Gaussian white noise for $\Delta{t}\geq20Min$.

\subsection{Gain/Loss Asymmetry}
\label{gain_loss}
\begin{figure}[H]
\centering
\begin{subfigure}[t]{.48\textwidth}
  \centering
  \includegraphics[width=1\linewidth]{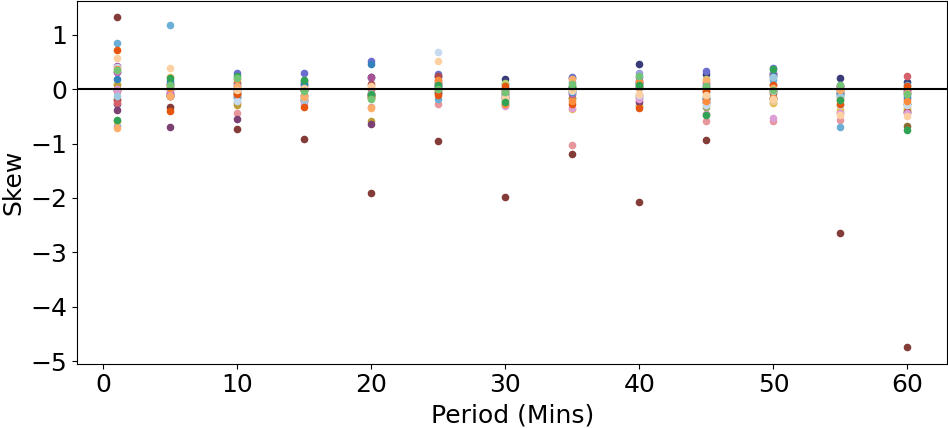}
\end{subfigure}
\begin{subfigure}[t]{.48\textwidth}
  \centering
  \includegraphics[width=1\linewidth]{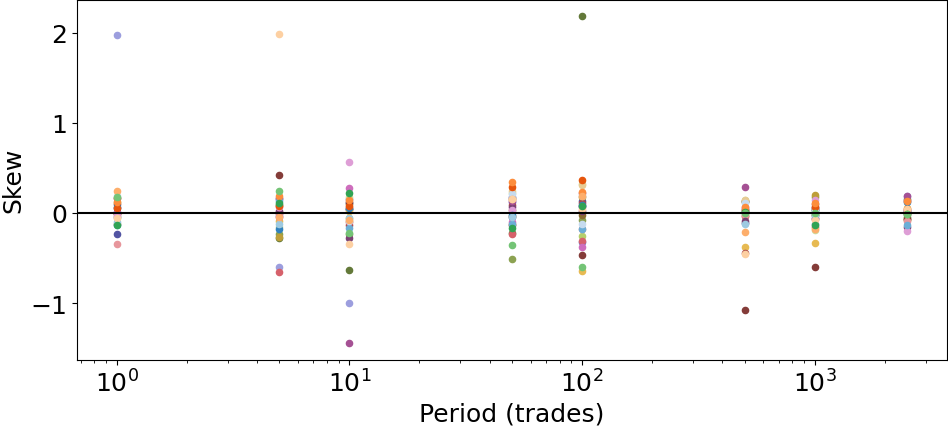}
\end{subfigure}
\caption{Skew of the return distributions in clock- and event-time.}
\label{fig:skew}
\end{figure}

Cont details his 3$^{rd}$ stylized fact (gain/loss asymmetry) as prices experiencing larger drawdowns than upward movements. As mentioned in Section \ref{literature}, it is not exactly clear what results were being referenced for this fact. Under some interpretations, returns should be expected to show negative skews, with skew measured as:
$$S\left(\Delta{t}\right)=\frac{\langle\left(r\left(t,\Delta{t}\right)-\langle r\left(t,\Delta{t}\right)\rangle\right)^3\rangle}{\sigma\left(\Delta{t}\right)^3}.$$
We do not see this property consistently across symbols for returns in clock- nor event-time, as shown in Fig. \ref{fig:skew}. For each timescale in either clock, we see a full range from negative to positive skewness for the return distributions.

We considered a more literal read of Cont's details for this fact as well, from which we would expect to see larger losses than we do gains. We measured the percentage of returns that are negative versus positive for different cutoffs. For a given timescale and quantile $q$, the cutoff is that quantile of a symbol's absolute returns in the timescale. The expectation is for most extreme returns to be losses and for this to be more true as the quantile-cutoff gets closer to the 100$^{th}$ percentile. We instead found the symbols split between having more extreme gains versus losses in each quantile and timescale\footnote{Shown in our supplementary material, Fig. \ref{fig:extreme_loss_perc}.}. We therefore do not find evidence of a gain/loss asymmetry effect in clock-time or event-time in our data.

\subsection{Volatility Clustering}
\label{vol_cluster}
Cont's 6$^{th}$ stylized fact expects volatility to cluster in time. We calculate volatility clustering by looking at the autocorrelation of absolute returns. Similar to linear ACF, let
$$C^0\left(\tau,\Delta{t}\right)=\text{corr}\left(|r\left(t,\Delta{t}\right)|, |r\left(t+\tau,\Delta{t}\right)|\right).$$
The expectation is for $C^0\left(\tau, \Delta{t}\right)>|C\left(\tau,\Delta{t}\right)|$ and for $C^0\left(\tau,\Delta{t}\right)$ to asymptotically go to zero, with a decay that looks roughly linear on a log-log plot. This latter property, a power-law decay of autocorrelation, is claimed in the details of Fact \#8.

\begin{figure}[H]
\centering
\begin{subfigure}[t]{.48\textwidth}
  \centering
  \includegraphics[width=1\linewidth]{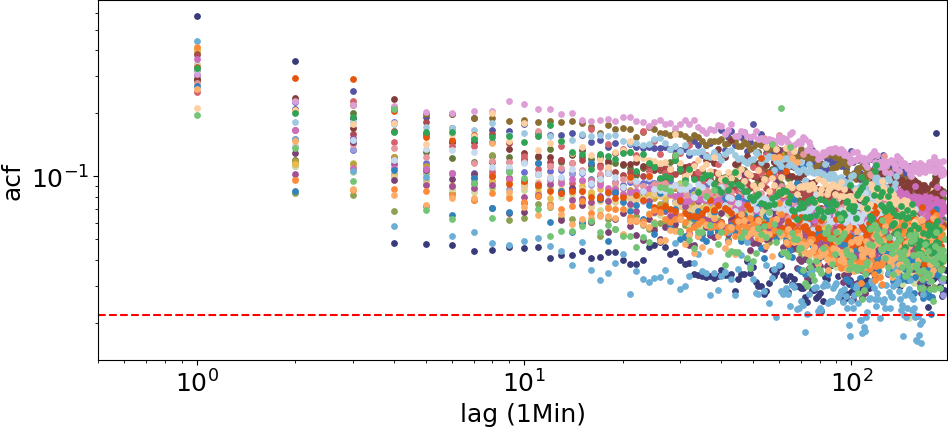}
\end{subfigure}
\begin{subfigure}[t]{.48\textwidth}
  \centering
  \includegraphics[width=1\linewidth]{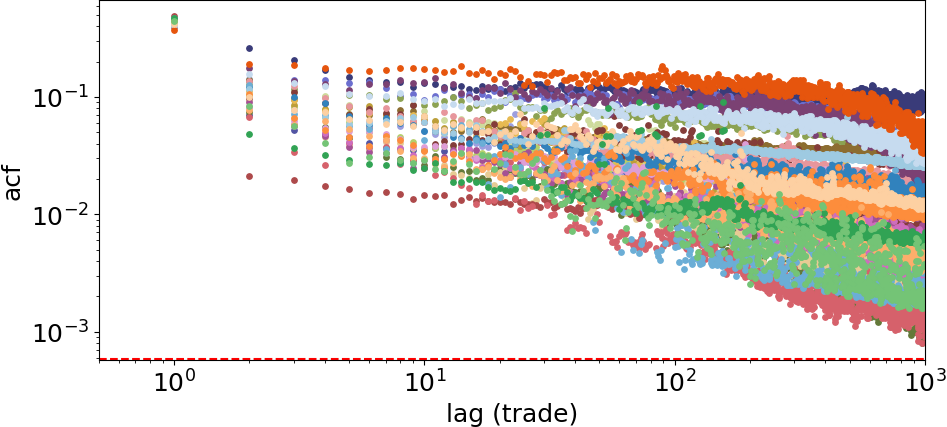}
\end{subfigure}
\caption{Autocorrelation of absolute returns.}
\label{fig:abs_autocorrelation}
\end{figure}

As shown in Fig. \ref{fig:abs_autocorrelation}, absolute ACF of 1-minute returns starts above the values we saw for linear autocorrelation (Section \ref{acf}) and for most symbols remain consistently above the range of white noise for the more than 100 lags tested. In log-log time, we see the decay asymptotically approaching zero. A couple symbols have absolute ACF near the levels of white noise by the last lags. We see similarly in trade-time, with the absolute ACF staring above 0.1 with a subsequent slow decay in log-log time. The rate of decay appears to slow after the first few lags and level out. An exponential cutoff appears to occur for some stocks at differing places in the later lags. The overall path suggests a slower than exponential decay, or `long-memory' of volatility in both timescales. The possible cutoffs to the effect seen in the later lags are worth further exploration, however, to gain confidence in the exact shape of the decay exhibited (power-law or some alternative).

\subsection{Intermittency}
\label{intermittency}

As detailed in Section \ref{literature}, the property of intermittency follows from and is intimately tied to volatility clustering and heavy tails in the literature. We attempt one additional measure of intermittency, however, through examining the distribution of interarrival times of extreme price moves. Specifically, consider the 99$^{th}$-percentile biggest absolute returns for a given symbol, which we will denote $N_{0.99,\Delta{t}}$. We can then count the number of these returns we see in a given period of time, with greater variability providing a measure of intermittency. We show these extreme returns occur more variably than they would if arising from a Poisson distribution by measuring their Fano factor. The Fano factor is defined as $$F\left(\Delta{t}\right)=\frac{\sigma^2_{N_{0.99,\Delta{t}}}}{\langle N_{0.99,\Delta{t}}\rangle},$$
the ratio of the variance to the mean for the number of extreme returns in a period. This ratio would be approximately one for a Poisson distribution, but we see this is not the case in Table \ref{table:ext_intermittency}. The Fano factor is greater than one for extreme trade-level returns in 1000 trades as well as the extreme 1-minute returns in 30-minute periods. We furthermore found the distribution of interarrival times between intraday extreme returns to show excess kurtosis\footnote{Shown in our supplementary material, Table \ref{table:ext_interarrival}}. From these findings, along with the heavy-tails and volatility clustering already discussed, we see evidence of intermittency in clock-time and event-time.

\begin{table}[H]
\centering
\begin{tabular}{|l|r|r|}
\hline
Symbol &  1-min Returns &  Trade Return \\
\hline
  AAPL &           2.69 &         52.12 \\
   AXP &           3.18 &         45.32 \\
    BA &           2.80 &         57.16 \\
   CAT &           2.87 &         48.14 \\
  CSCO &           2.82 &         31.67 \\
   CVX &           2.90 &         42.68 \\
   DIS &           2.59 &         51.27 \\
  DWDP &           2.88 &         44.43 \\
    GS &           3.21 &         40.77 \\
    HD &           2.83 &         50.90 \\
   IBM &           2.72 &         48.49 \\
  INTC &           3.00 &         26.95 \\
   JNJ &           4.70 &         52.49 \\
   JPM &           3.23 &         41.07 \\
    KO &           3.36 &         27.61 \\
   MCD &           3.07 &         57.47 \\
   MMM &           2.74 &         53.67 \\
   MRK &           3.17 &         45.73 \\
  MSFT &           3.24 &         48.15 \\
   NKE &           3.70 &         38.30 \\
   PFE &           3.02 &         32.33 \\
    PG &           3.23 &         45.85 \\
   TRV &           3.16 &         42.26 \\
   UNH &           2.76 &         53.56 \\
   UTX &           3.72 &         58.03 \\
     V &           3.39 &         60.34 \\
    VZ &           2.65 &         37.01 \\
   WBA &           3.34 &         45.83 \\
   WMT &           3.84 &         53.78 \\
   XOM &           2.67 &         34.32 \\
\hline
\end{tabular}
\caption{Fano factor of extreme returns. For returns with magnitudes in the 99$^{th}$-quantile, we measured the Fano factor of the number of extreme returns in a coarse period of time. For 1-minute returns, the coarse period was 30-minutes, for trade-level returns, the coarse period was 1000 trades.}
\label{table:ext_intermittency}
\end{table}

\subsection{Leverage Effect}
\label{leverage}
\begin{figure}[H]
\centering
\begin{subfigure}[t]{.48\textwidth}
    \centering
    \includegraphics[width=1\linewidth]{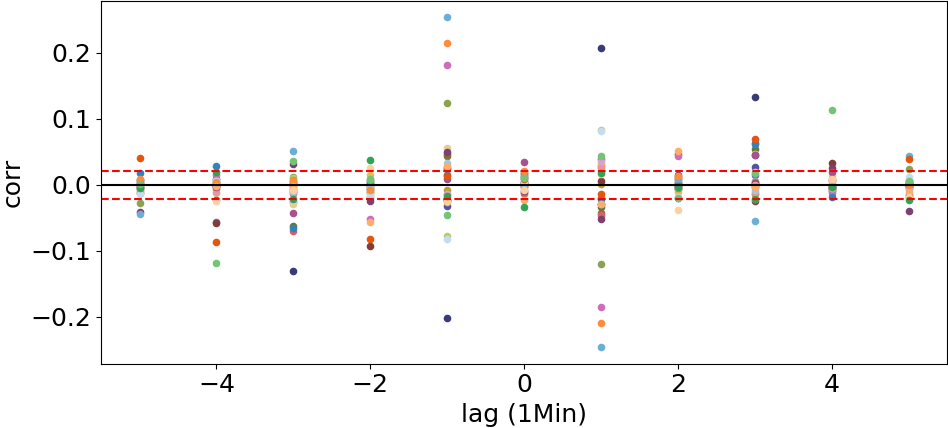}
\end{subfigure}%
\hspace{1em}
\begin{subfigure}[t]{.48\textwidth}
    \centering
    \includegraphics[width=1\linewidth]{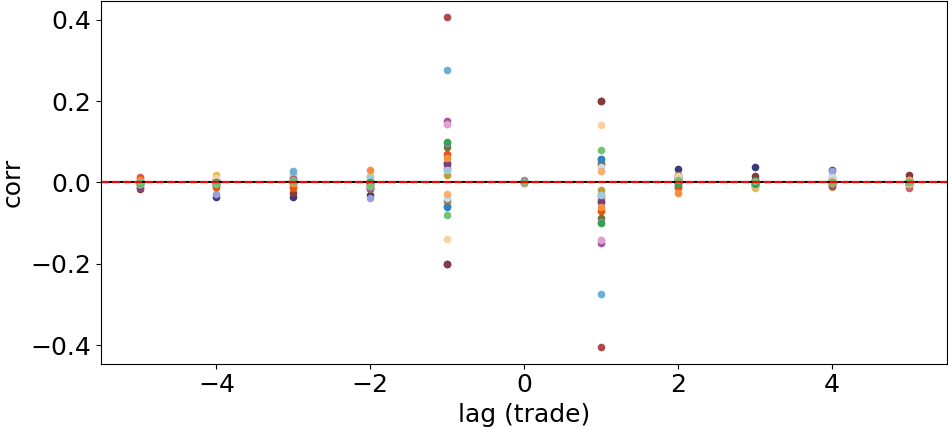}
\end{subfigure}%
\caption{Leverage effect, measured as the correlation between $r\left(t,\Delta{t}\right)$ and $|r\left(t+\tau,\Delta{t}\right)|$.}
\label{fig:leverage_effect}
\end{figure}

Cont's 9$^{th}$ stylized fact asserts that volatility is negatively correlated with the returns for an asset. We measure the leverage effect (Fact \#9) as Cont laid out in \cite{cont_empirical_2001}, drawing from the results of Bouchaud et al. \cite{bouchaud_leverage_2001} and Pagan \cite{pagan_econometrics_1996}. This approach looks simply at the correlation between returns and lagged volatility, with volatility measured as the absolute returns\footnote{We also looked at the squared returns for the volatility, as used by Bouchaud et al. \cite{bouchaud_leverage_2001}. There was no material effect to our results, as shown in our supplemental material, and we therefore use the absolute returns to be consistent with the rest of our analysis.}: 
$$L\left(\tau, \Delta{t}\right) = \text{corr}\left(r\left(t,\Delta\left(t\right)\right),|r\left(t+\tau,\Delta{t}\right)|\right).$$
The expectation is for $L\left(\tau\right)<0$ when $\tau=1$ and for $L\left(\tau\right)<L\left(-\tau\right)$ when $\tau>0$.

We find no clear trend to the correlation values across symbols. There are varying strengths and signs to the correlations at each lag, suggesting they might arise from specific variation in the path of a given symbol's price over our observation period. For some symbols, there is an interesting symmetry where $L\left(-1\right)\approx -L\left(1\right)$. This also goes against the descriptions from Cont \cite{cont_empirical_2001} and Bouchaud et al. \cite{bouchaud_leverage_2001}, who described the relationship between returns and negatively lagged volatility as being largely insignificant. Overall, we do not see the expected direction of the leverage effect relationship, nor do we see any clear trend in this relationship across the symbols.

\subsection{Volume/Volatility Correlation}
\label{volume_vol}
\begin{figure}[H]
\centering
\begin{subfigure}[t]{.48\textwidth}
    \centering
    \includegraphics[width=1\linewidth]{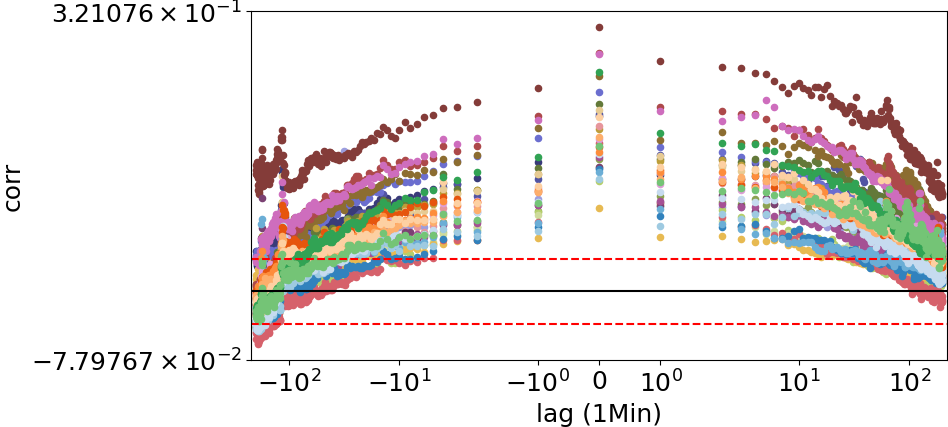}
\end{subfigure}
\hspace{1em}
\begin{subfigure}[t]{.48\textwidth}
    \centering
    \includegraphics[width=1\linewidth]{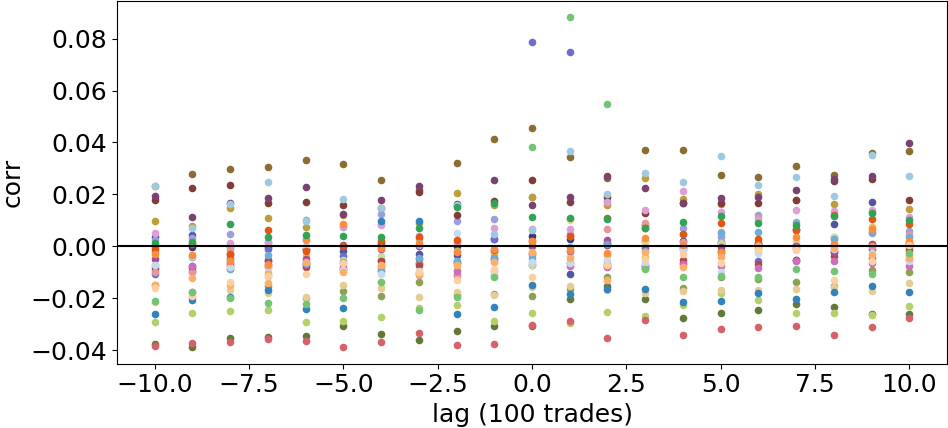}
\end{subfigure}%
\caption{Share volume versus lagged volatility.}
\label{fig:volume_volatility}
\end{figure}

The relationship between volume and volatility is Cont's 10$^{th}$ stylized fact and asserts a positive correlation between volume and volatility. We examine this by taking the correlations of the volume of shares\footnote{In clock-time, correlation between the volume of trades and volatility was very similar to that between shares and volatility. We thus use volume of shares here to have a consistent measure between clock- and event-time.} in a period with the lagged absolute returns. Indeed, in clock-time we do see a strong, positive correlation between shares traded and volatility for each of the symbols. There is a similar pattern to the decay of this correlation across the symbols, but some go below the range of white noise due to a lower starting correlation at lag zero. This relationship goes away in event-time. We looked at $N=1$ and $N=100$ in event-time, finding no clear trend across symbols in either timescale. The correlation is weak and varies in sign depending on the symbol, in contrast to the clock-time results. These results give some explanation towards the differences in how trade-level and clock-time returns behave\footnote{Indeed, we see a slow decay in the autocorrelation of $Vol(t,\Delta{t})$ in clock-time, similar to the volatility clustering discussed in Facts \#6 and 8. This is given in our supplemental materials.}.

\subsection{Asymmetry in Timescales}
\label{asymmetry}
The final stylized fact (Fact \#11) examines the asymmetry of the flow of information across timescales. For our base timescales $\Delta{t}$ of 1-minute, we consider the coarse-grained timescale $\Delta{T}=30Min$. In event-time, we use a base timescale of trade-time and $N=1000$ as our coarse timescale $\Delta{T}$. We calculate $A\left(\tau,\Delta{t},\Delta{T}\right)=\text{corr}\left(\overline{|r\left(t\in T,\Delta{T}\right)|}, |r\left(T+\tau,\Delta{T}\right)|\right),$ as done by M\"uller et al. \cite{muller_volatilities_1997}. We measure the asymmetry by differencing the correlation at the corresponding positive and negative lags $\tau$:
$$D\left(\tau,\Delta{t},\Delta{T}\right)=A\left(\tau,\Delta{t},\Delta{T}\right)-A\left(-\tau,\Delta{t},\Delta{T}\right).$$ The expectation is for negative lags to have a larger correlation than the corresponding positive lags for at least a few steps.

\begin{figure}[H]
\centering
\begin{subfigure}[t]{.48\textwidth}
    \centering
    \includegraphics[width=1\linewidth]{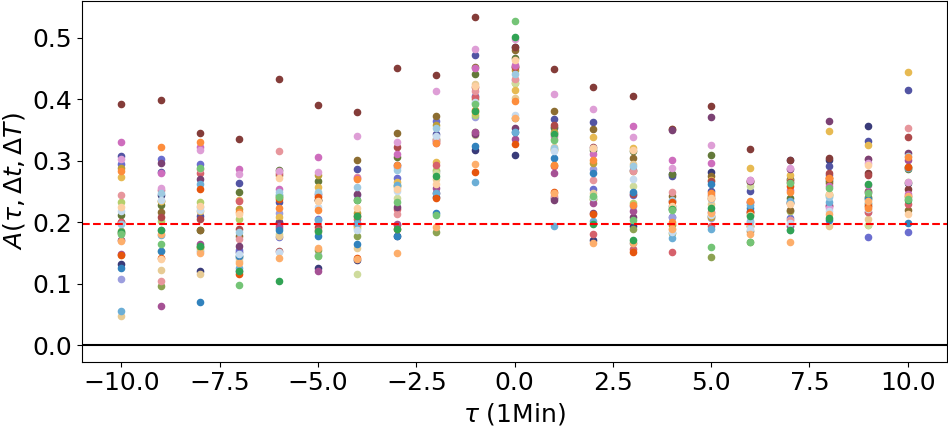}
    \caption{1-minute versus lagged 30-minute volatility.}
    \label{fig:asymmetry_1min_30min}
\end{subfigure}%
\hspace{1em}
\begin{subfigure}[t]{.48\textwidth}
    \centering
    \includegraphics[width=1\linewidth]{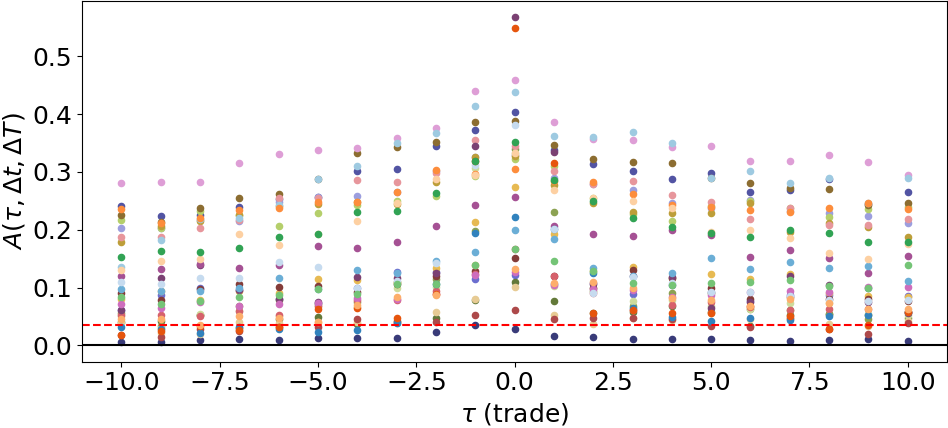}
    \caption{Trade versus lagged 1000-trade volatility.}
    \label{fig:asymmetry_trade_1000}
\end{subfigure}%
\caption{Average fine volatility versus lagged coarse-grained volatility.}
\label{fig:asymmetry_timescales}
\end{figure}

\begin{figure}[H]
\centering
\begin{subfigure}[t]{.48\textwidth}
    \centering
    \includegraphics[width=1\linewidth]{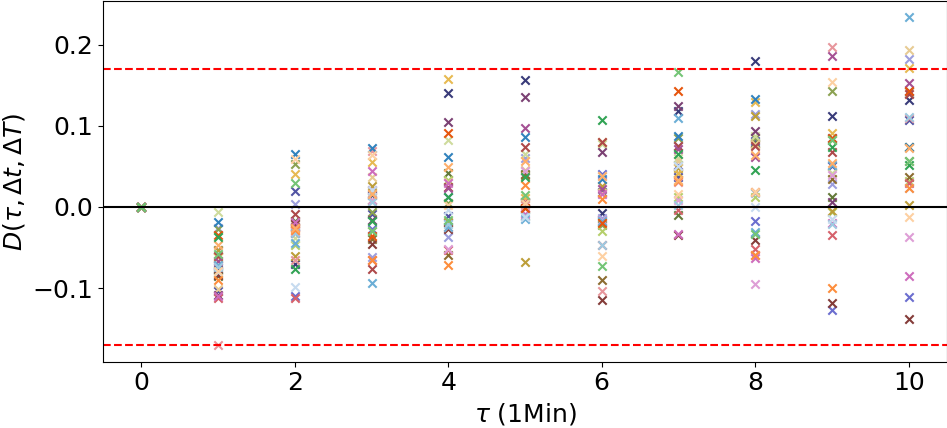}
    \caption{1-minute versus lagged 30-minute volatility.}
    \label{fig:asymmetry_1min_30min_diff}
\end{subfigure}%
\hspace{1em}
\begin{subfigure}[t]{.48\textwidth}
    \centering
    \includegraphics[width=1\linewidth]{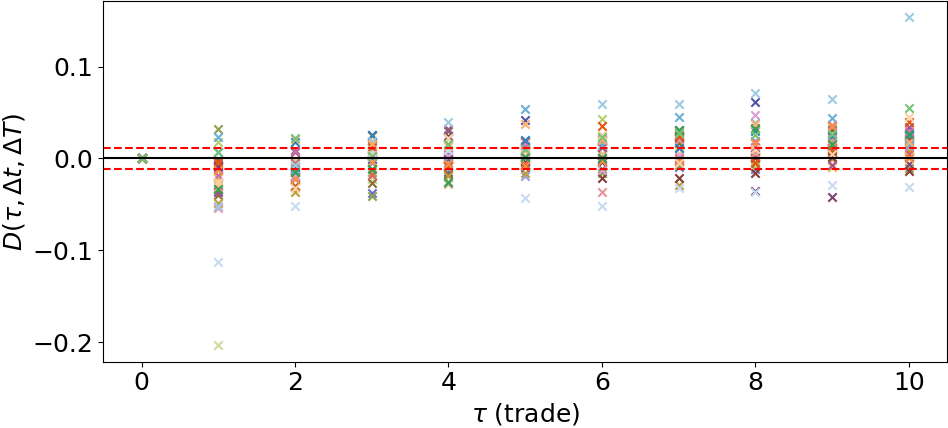}
    \caption{Trade versus lagged 1000-trade volatility.}
    \label{fig:asymmetry_trade_1000_diff}
\end{subfigure}%
\caption{Asymmetry effect --- Difference between positive and negative lags for the correlation between the average fine volatility and lagged coarse volatility.}
\label{fig:asymmetry_timescales_diff}
\end{figure}

We first of all see a positive relationship between coarse-grained volatility and fine-grained volatility, synchronously and at a lag, for each timescale and perspective (Fig. \ref{fig:asymmetry_timescales}). As in M\"uller et al. \cite{muller_volatilities_1997}, the relationship is strongest at lag $\tau=0$, particularly in clock-time. In event-time, the relationship is more varied, and some symbols have a correlation near zero for all lags. Comparing positive to negative lags gives an indication of a possible causal relationship. There is no clear trend in the difference $D\left(\tau,\Delta{t},\Delta{T}\right)$ across symbols, however (Fig. \ref{fig:asymmetry_timescales_diff}). We would expect $D\left(\tau,\Delta{t},\Delta{T}\right)<0$ for at least a few lags. In clock-time, the first-lag difference $D\left(\tau=1,\Delta{t}=1Min,\Delta{T}=30Min\right)$ is negative for each symbol, but some values are nearly zero. Past the first lag, symbols are split between negative and positive differences for each $\tau$. The range of values seen on the random white noise returns is also larger for this effect than we saw for other measures, and the negative differences fall within this range for the first lag. In event-time, the differences were split between positive and negative for every lag.
It is possible the asymmetry effect is more plausible when considered for symbols on average, at least in clock-time. We discuss this possibility more in the next section. However, for individual symbols we do not find a consistent negative difference to indicate a clear asymmetry of volatility prediction across timescales.

\section{Discussion}
\label{discussion}
As summarized in Table \ref{table:fact_summary}, we have shown evidence for eight of Cont's 11 stylized facts holding true for intraday returns for each of the Dow 30 stocks in our sample. How we constructed time in our analysis mattered in how (or whether) the facts manifested. We found positive correlation between volume and volatility (Fact \#10) in clock-time but not in general for the trade-level returns. Meanwhile, aggregational Gaussianity (Fact \#4) was more clearly expressed in event-time than clock-time, although the property is overall expressed in both clocks.

The measures examined for Facts \#3, \#9, and \#11 do not provide evidence of gain/loss asymmetry, the leverage effect, or asymmetry in timescales, respectively, being present in general for intraday returns of individual stocks. Instead, the measures were found to vary in sign and strength depending on the symbol and timescale. Alternative approaches looking at longer timescales (e.g. daily) or grouping stocks together might yield different results for these facts. An example is the analysis done by Chicheportiche and Bouchaud \cite{chicheportiche_fine-structure_2014}, who looked at daily data from 280 stocks from 2000-2009. They standardized the returns such that they could measure the leverage effect and asymmetry in timescales for all stocks together, finding evidence of both effects. Blanc et al. \cite{blanc_quadratic_2017} took a similar approach using 5-minute returns of 133 stocks from 2000-2009. They found evidence of asymmetry in timescales but noted that the leverage effect is insignificant for intraday returns. Looking at the motion of multiple stocks together (such as an index or the average price over multiple stocks) is another approach taken in the literature. Bouchaud et al. \cite{bouchaud_leverage_2001} found the average leverage effect for daily index prices was stronger than for daily stock prices.
Research measuring gain/loss asymmetry through investment horizons found the effect to appear for indices but not for individual stocks \cite{jensen_inverse_2003}\cite{johansen_optimal_2006}\cite{siven_multiscale_2009}. Future research could therefore examine the full set of Cont's stylized facts for groups of stocks collectively or using daily prices.

Our results provide an improved understanding of how Cont's stylized facts should be interpreted in modern markets. In particular, eight of the stylized facts appear to be well-supported for intraday returns of individual stocks, while the remaining three were not present for the measures examined. This has implications for our understanding of stock markets and how we should evaluate models thereof. Most agent-based models of markets evaluate their models on the simulated price series, which is frequently more analogous to event-time in the empirical market than to clock-time \cite{chakraborti_econophysics_2011}. The developer of such a model can take confidence in using at least seven of Cont's stylized facts to evaluate their model's relevance to the empirical market, while caution should be taken if expecting a strong volume/volatility relationship depending on how time is being evaluated. The remaining three facts may be more appropriate in evaluating a model with multiple assets traded simultaneously and with the price sampled infrequently.

Finally, Cont's stylized facts are not exhaustive in capturing characteristics of financial markets. Examining the commonality of other features of financial markets such as daily volatility patterns \cite{andersen_intraday_1997}\cite{wood_investigation_1985}, extreme price events \cite{johnson_abrupt_2013}\cite{mcgroarty_high_2019}, and order book characteristics \cite{gould_limit_2013} can help widen our understanding of stylized facts and provide additional benchmarks for market models to hit. 

\section{Conclusion}
\label{conclusion}
Cont’s original set of stylized facts \cite{cont_empirical_2001} emerged from a synthesis of empirical studies, each study focused on a market which existed prior to 2001. As demonstrated elsewhere in previous studies, the technological arms race and resulting market fragmentation in the intervening decades since Cont’s study fundamentally changed the dynamics of the U.S. stock market \cite{johnson_abrupt_2013}\cite{tivnan_fragmentation_2020}\cite{van_oort_ecological_2022}. Motivated by these market changes to revisit Cont’s original study, we find strong evidence for eight of Cont’s original set of 11 stylized facts when looking at intraday returns of individual stocks. 
A robust set of stylized facts serves at least two distinct communities. For the community of financial regulators, the set of stylized facts provides guideposts against which to assess the impacts of regulatory reform, both the intended and unintended impacts.  For the scientific community, the set of stylized facts provides the guideposts for the design, development, test and calibration for the next generation of market models.

\section*{Acknowledgment}

The authors gratefully acknowledge helpful discussions with Anshul Anand, James Angel, Chris Bassler, Lashon Booker, Jean-Philippe Bouchaud, Robert Brooks, Eric Budish, Richard Byrne, Peter Carrigan, Christopher Danforth, Matthew Dinger, Peter Sheridan Dodds, Andre Frank, Bill Gibson, Frank Hatheway, Emily Hiner, Chuck Howell, Eric Hunsader, Robert Jackson, Neil Johnson, Will Kirkman, Michael Kometer, Blake LeBaron, Phil Mackintosh, Matthew McMahon, Beth Meinert, Matthew Mihalcin, Rishi Narang, Mark Phillips, Joseph Saluzzi, Brendan Tivnan, Kevin Toner, Jason Veneman, and Thomas Wilk. All opinions and remaining errors are the sole responsibility of the authors and do not reflect the opinions nor perspectives of their affiliated institutions nor that of the funding agencies. The authors declare no conflicts of interest.

\end{sloppypar}
\bibliographystyle{plainnat}
\bibliography{refs}

\appendix
\section{Appendix}

\subsection{Results on normalized returns}
\label{normailized}
As noted in Section \ref{cont_facts}, there is a U-shaped daily volatility pattern characteristic to financial markets. This has possible implications for statistical analysis on returns and volatility and could possibly obscure our findings. 
We therefore considered normalized returns in addition to raw returns calculated on the timescales already discussed. Specifically, we first normalized the returns from each day $T$ by the standard deviation for that day at the given timescale, $\sigma\left(T, \Delta{t}\right)$:
$$r'_T\left(t,\Delta{t}\right)=\frac{r\left(t,\Delta{t}\right)}{\sigma\left(T, \Delta{t}\right)}$$
Next, for a clock-timescale $\Delta{t}$, let
$$v\left(t,\Delta{t}\right)=\langle r'_T\left(t,\Delta{t}\right)\rangle_{T=1,2,...103},$$
the average absolute return at time $t$ across the 103-day sample for the stock. Let 
$$\hat{r}_T\left(t,\Delta{t}\right)=\frac{r'_T\left(t,\Delta{t}\right)}{v\left(t,\Delta{t}\right)}.$$ 
We did not do this latter normalization in event-time, since event-time is already correcting for different activity levels at different times of day by fixing the number of trades in a bucket. In other words, for an event-timescale $N$, $\hat{r}$ is simply:
$$\hat{r}_T\left(t,N\right)=r'_T\left(t,N\right).$$
This gives us two different views of normalized returns in addition to the raw returns calculated in either timescale. Please note, this is the same normalization technique used in Section \ref{heavy_tails} to examine Fact \#7, conditional heavy tails.

\begin{figure}[H]
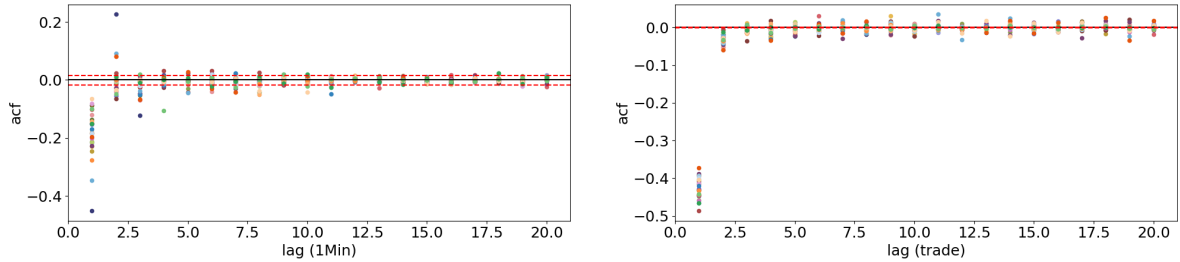

\centering
\begin{subfigure}[t]{.48\textwidth}
  \centering
  \includegraphics[width=1\linewidth]{1Min_sips_dow__D_Xautocorrelation_of_returns_summary.png}
\end{subfigure}
\hspace{1em}
\begin{subfigure}[t]{.48\textwidth}
  \centering
  \includegraphics[width=1\linewidth]{trade_sips_dow__D_Xautocorrelation_of_returns_summary.png}
\end{subfigure}
\caption{Linear autocorrelation of returns --- Raw.}
\label{fig:autocorrelation_raw}
\end{figure}

\begin{figure}[H]
\centering
\begin{subfigure}[t]{.48\textwidth}
  \centering
  \includegraphics[width=1\linewidth]{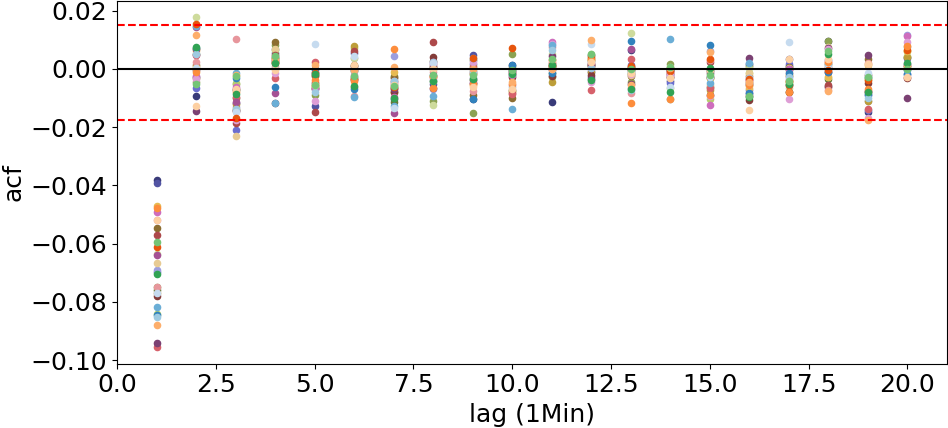}
\end{subfigure}
\hspace{1em}
\begin{subfigure}[t]{.48\textwidth}
  \centering
  \includegraphics[width=1\linewidth]{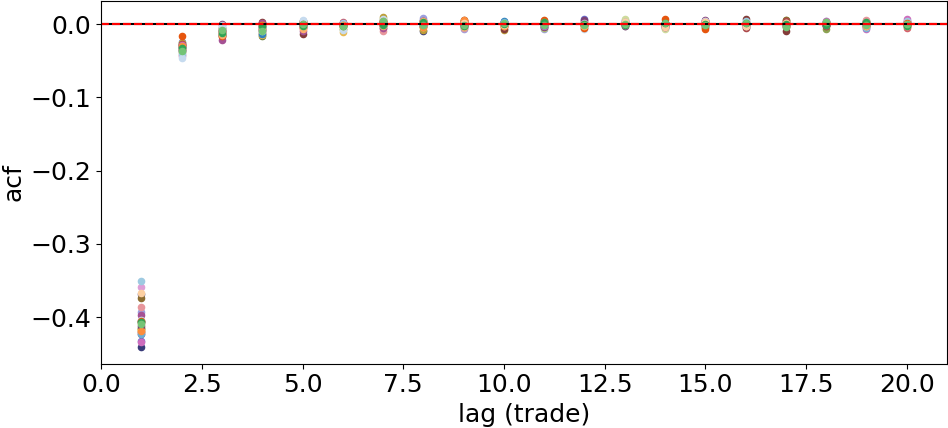}
\end{subfigure}
\caption{Linear autocorrelation of returns --- Normalized.}
\label{fig:autocorrelation_d_norm}
\end{figure}

\begin{figure}[H]
\centering
\begin{subfigure}[t]{.48\textwidth}
  \centering
  \includegraphics[width=1\linewidth]{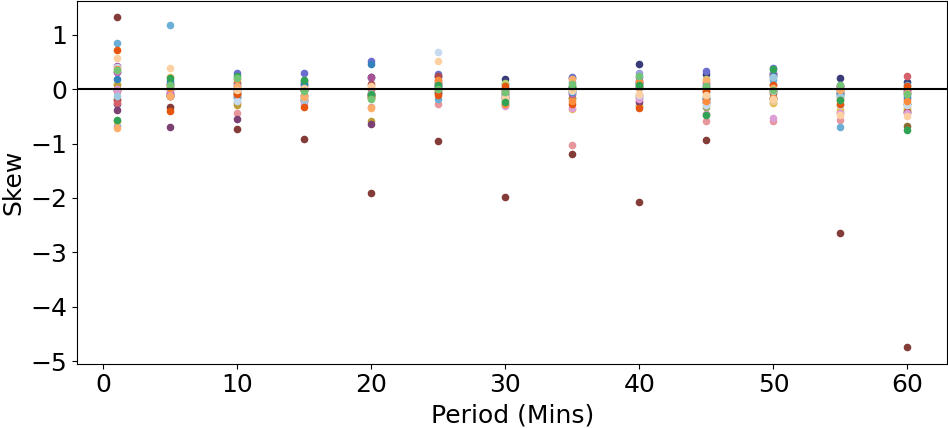}
\end{subfigure}
\begin{subfigure}[t]{.48\textwidth}
  \centering
  \includegraphics[width=1\linewidth]{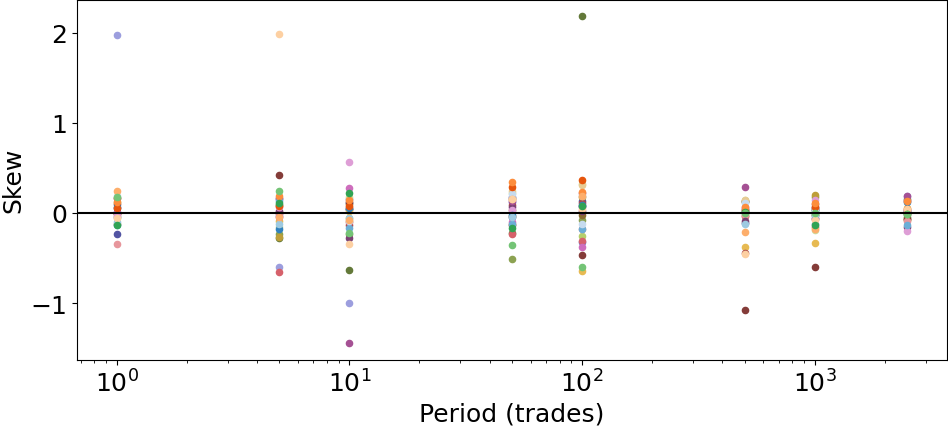}
\end{subfigure}
\caption{Skew of the return distributions in clock- and event-time --- Raw.}
\label{fig:skew_raw}
\end{figure}

\begin{figure}[H]
\centering
\begin{subfigure}[t]{.48\textwidth}
  \centering
  \includegraphics[width=1\linewidth]{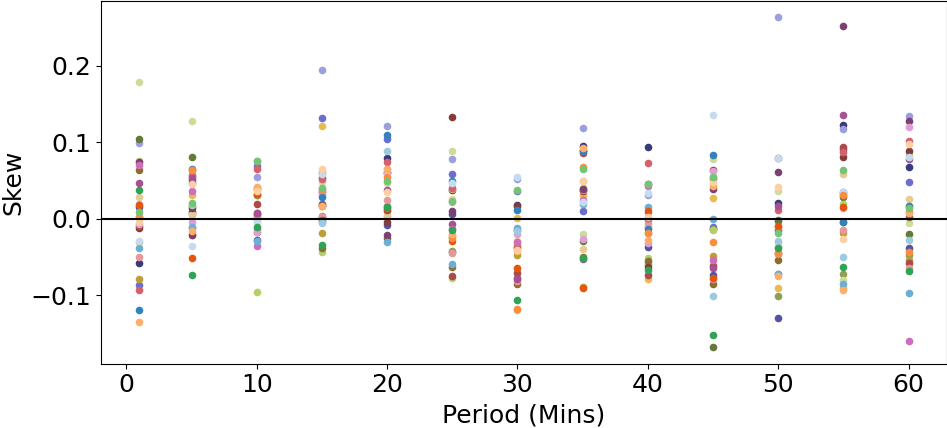}
\end{subfigure}
\begin{subfigure}[t]{.48\textwidth}
  \centering
  \includegraphics[width=1\linewidth]{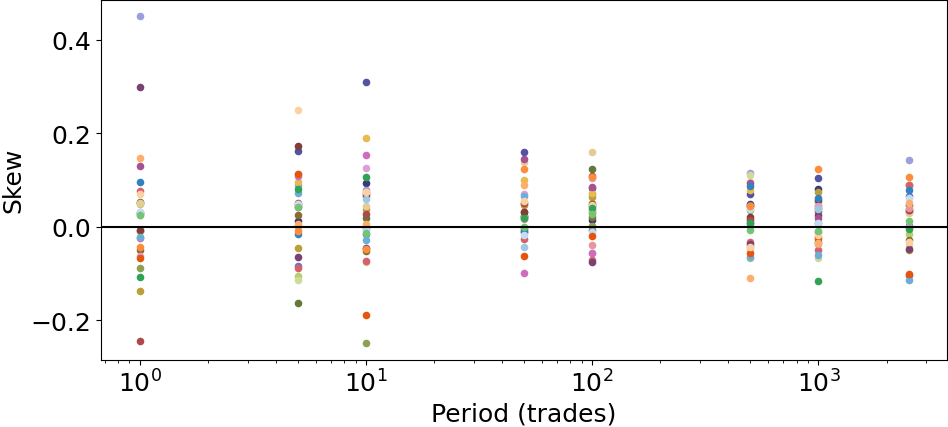}
\end{subfigure}
\caption{Skew of the return distributions in clock- and event-time --- Normalized.}
\label{fig:skew_norm}
\end{figure}

\begin{figure}[H]
\centering
\begin{subfigure}[t]{.48\textwidth}
  \centering
  \includegraphics[width=1\linewidth]{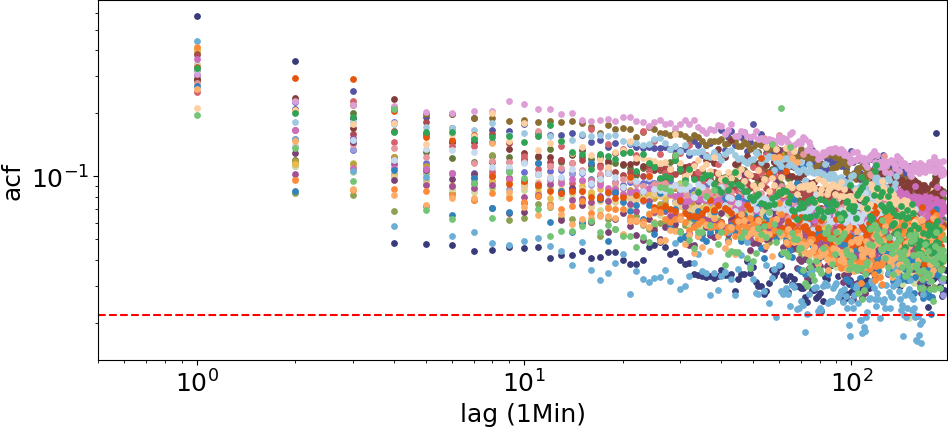}
\end{subfigure}
\begin{subfigure}[t]{.48\textwidth}
  \centering
  \includegraphics[width=1\linewidth]{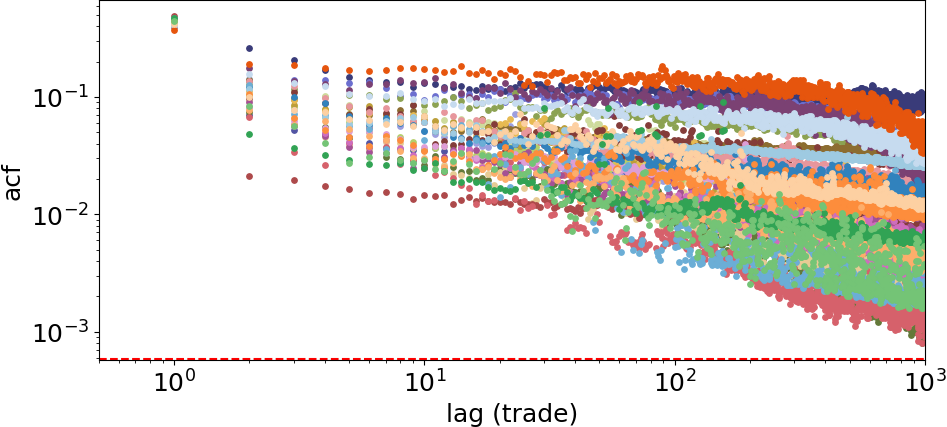}
\end{subfigure}
\caption{Autocorrelation of absolute returns --- Raw.}
\label{fig:abs_autocorrelation_raw}
\end{figure}

\begin{figure}[H]
\centering
\begin{subfigure}[t]{.48\textwidth}
  \centering
  \includegraphics[width=1\linewidth]{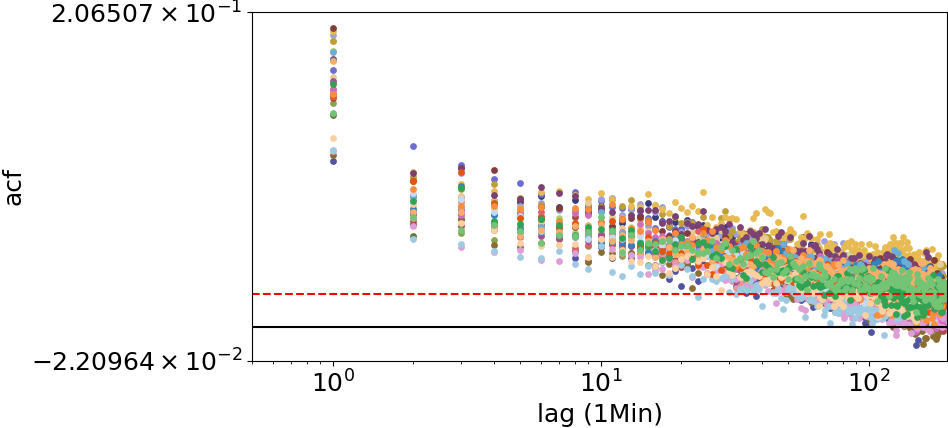}
\end{subfigure}
\begin{subfigure}[t]{.48\textwidth}
  \centering
  \includegraphics[width=1\linewidth]{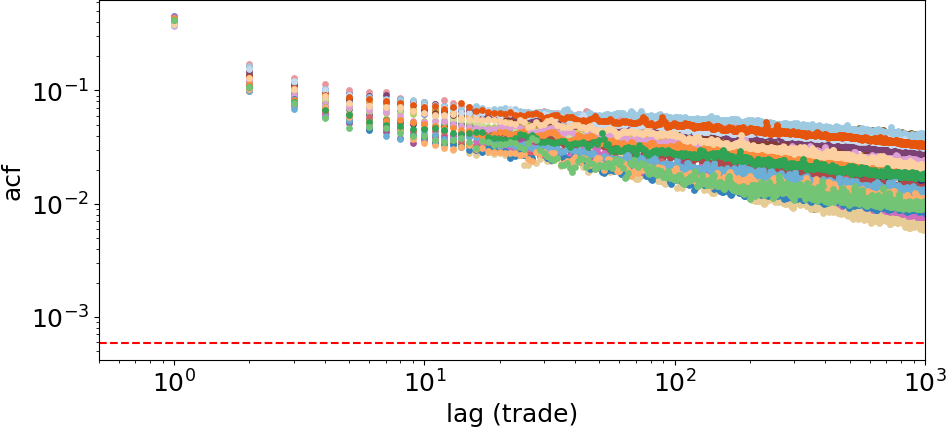}
\end{subfigure}
\caption{Autocorrelation of absolute returns --- Normalized.}
\label{fig:abs_autocorrelation_norm}
\end{figure}

\begin{figure}[H]
\centering
\begin{subfigure}[t]{.48\textwidth}
    \centering
    \includegraphics[width=1\linewidth]{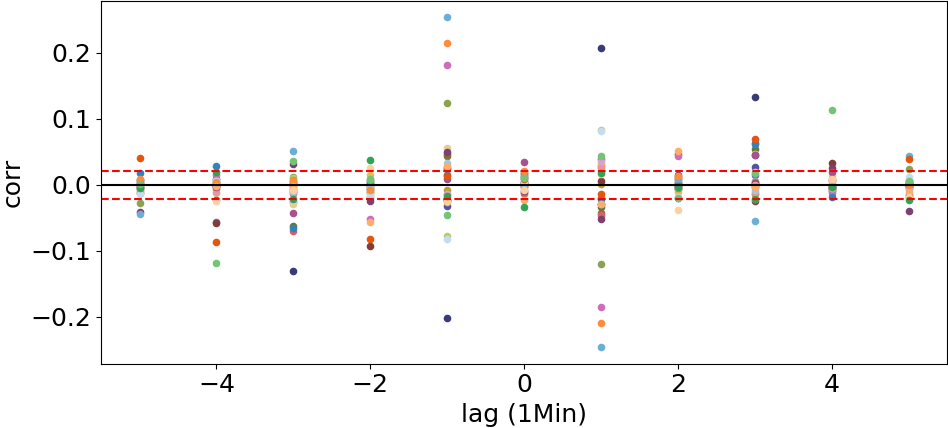}
\end{subfigure}%
\hspace{1em}
\begin{subfigure}[t]{.48\textwidth}
    \centering
    \includegraphics[width=1\linewidth]{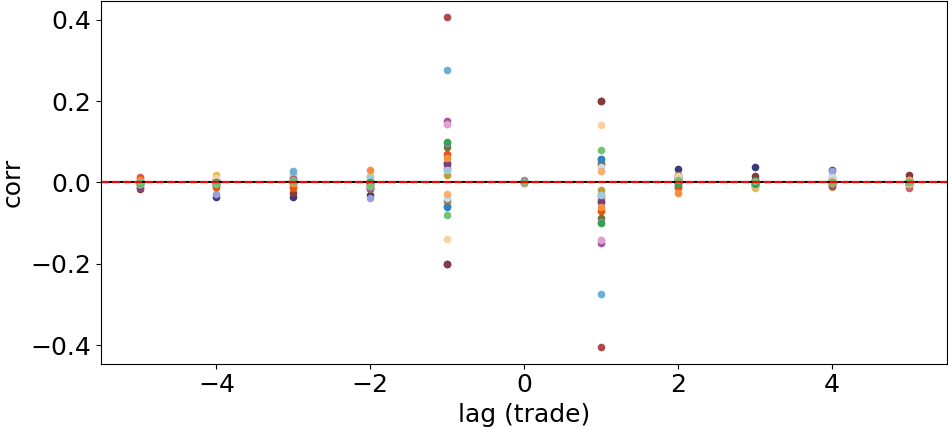}
\end{subfigure}%
\caption{Leverage effect, measured as the correlation between $r\left(t,\Delta{t}\right)$ and $|r\left(t+\tau,\Delta{t}\right)|$.}
\label{fig:leverage_effect_raw}
\end{figure}

\begin{figure}[H]
\centering
\begin{subfigure}[t]{.48\textwidth}
    \centering
    \includegraphics[width=1\linewidth]{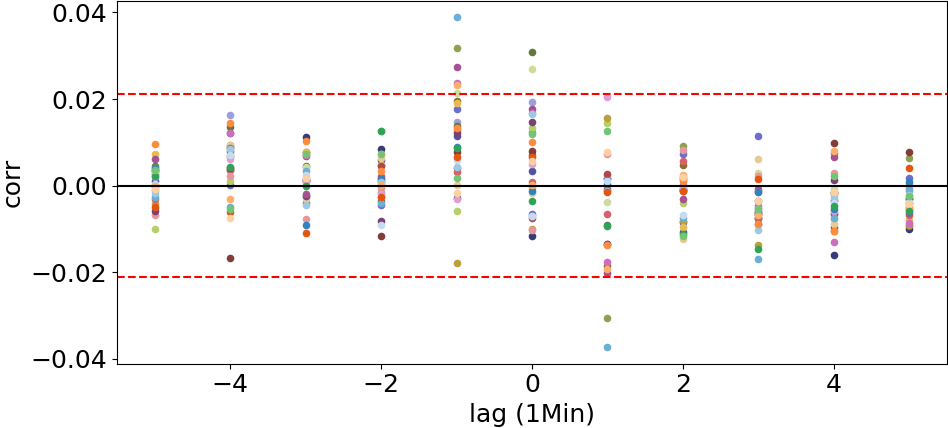}
\end{subfigure}%
\hspace{1em}
\begin{subfigure}[t]{.48\textwidth}
    \centering
    \includegraphics[width=1\linewidth]{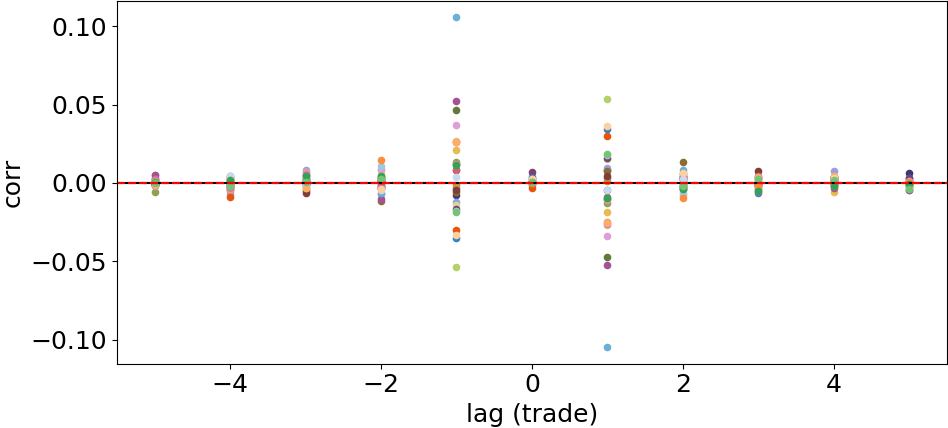}
\end{subfigure}%
\caption{Leverage effect, measured as the correlation between $r\left(t,\Delta{t}\right)$ and $|r\left(t+\tau,\Delta{t}\right)|$ --- Normalized.}
\label{fig:leverage_effect_norm}
\end{figure}

\begin{figure}[H]
\centering
\begin{subfigure}[t]{.48\textwidth}
    \centering
    \includegraphics[width=1\linewidth]{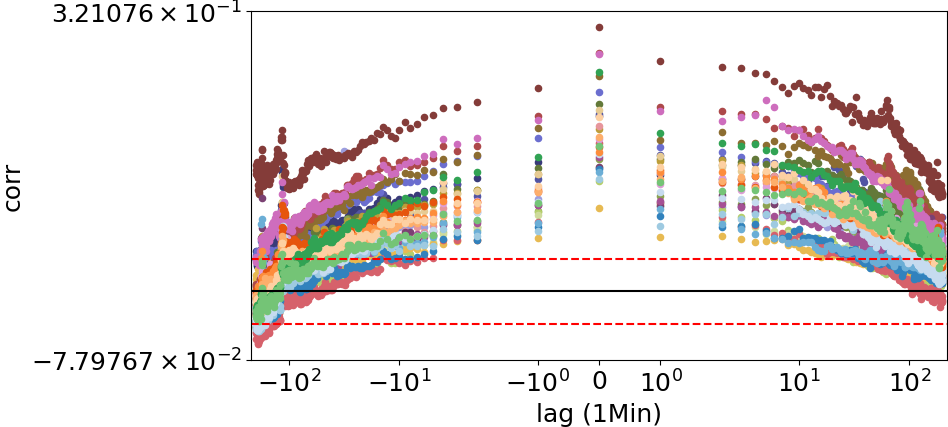}
\end{subfigure}
\hspace{1em}
\begin{subfigure}[t]{.48\textwidth}
    \centering
    \includegraphics[width=1\linewidth]{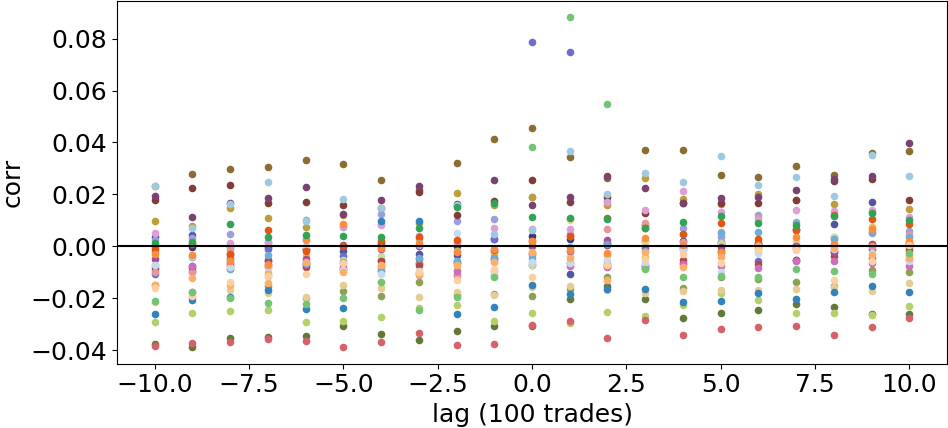}
\end{subfigure}%
\caption{Share volume versus lagged volatility --- Raw.}
\label{fig:volume_volatility_raw}
\end{figure}

\begin{figure}[H]
\centering
\begin{subfigure}[t]{.48\textwidth}
    \centering
    \includegraphics[width=1\linewidth]{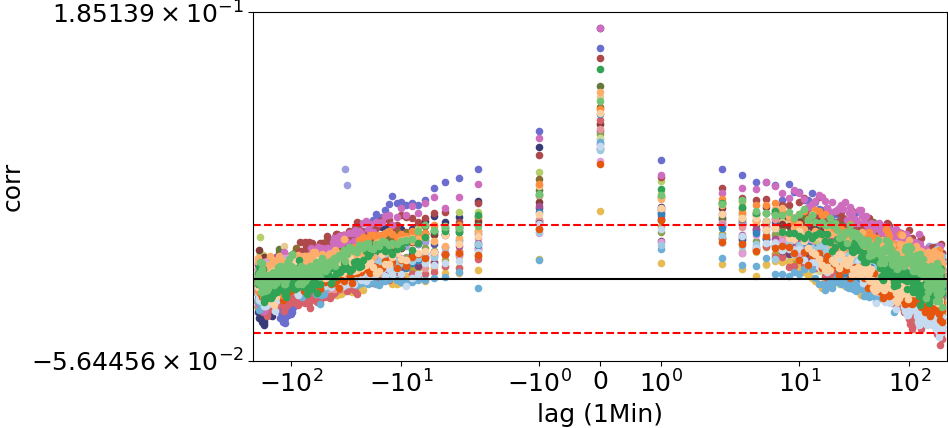}
\end{subfigure}
\hspace{1em}
\begin{subfigure}[t]{.48\textwidth}
    \centering
    \includegraphics[width=1\linewidth]{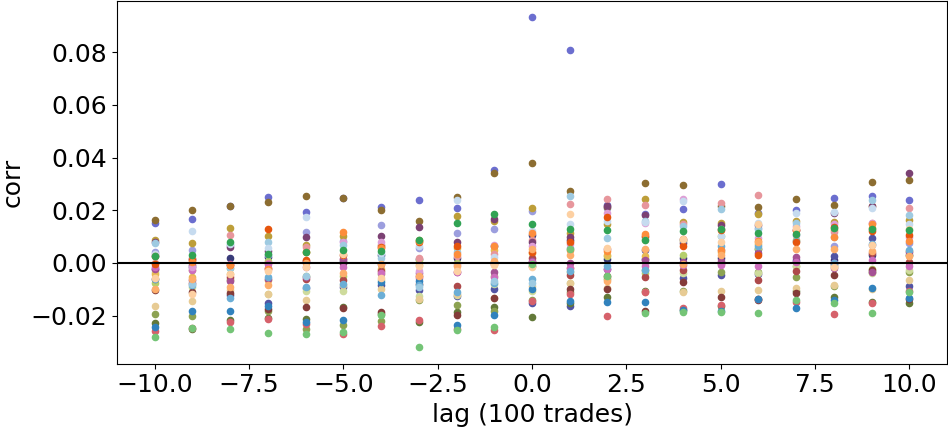}
\end{subfigure}%
\caption{Share volume versus lagged volatility --- Normalized.}
\label{fig:volume_volatility_norm}
\end{figure}

\begin{figure}[H]
\centering
\begin{subfigure}[t]{.48\textwidth}
    \centering
    \includegraphics[width=1\linewidth]{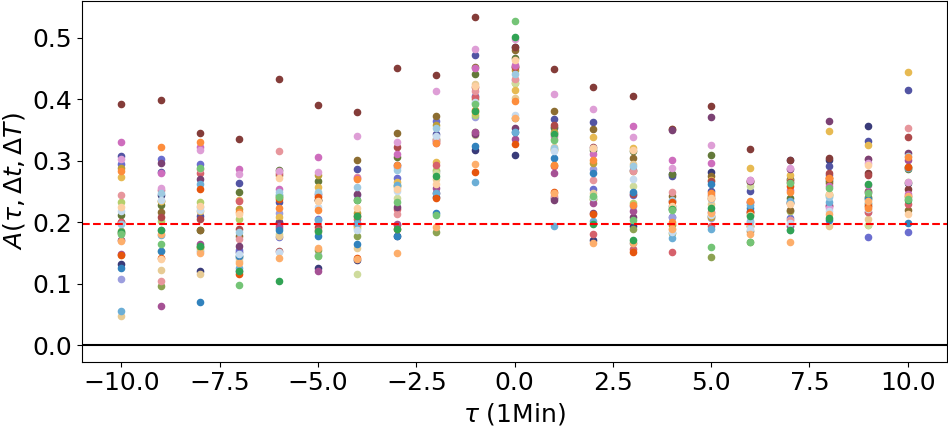}
    \caption{1-minute versus lagged 30-minute volatility.}
    \label{fig:asymmetry_1min_30min_raw}
\end{subfigure}%
\hspace{1em}
\begin{subfigure}[t]{.48\textwidth}
    \centering
    \includegraphics[width=1\linewidth]{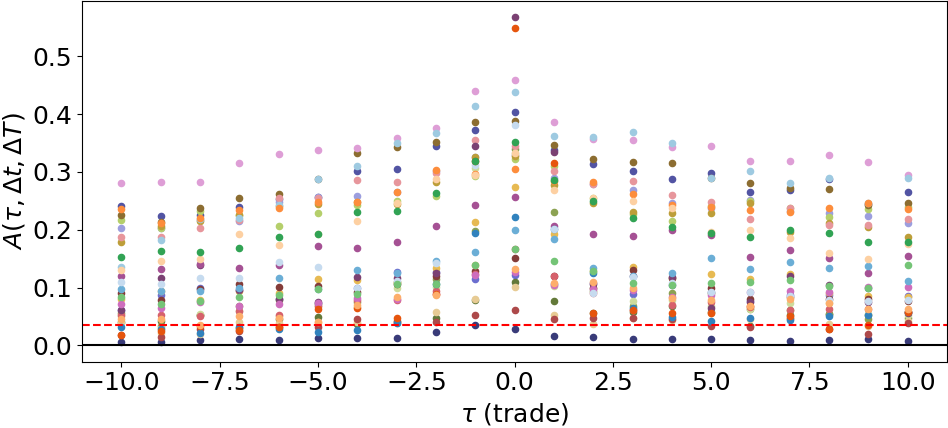}
    \caption{Trade versus lagged 1000-trade volatility.}
    \label{fig:asymmetry_trade_1000_raw}
\end{subfigure}%
\caption{Average fine-grained volatility versus lagged coarse-grained volatility --- Raw.}
\label{fig:asymmetry_timescales_raw}
\end{figure}

\begin{figure}[H]
\centering
\begin{subfigure}[t]{.48\textwidth}
    \centering
    \includegraphics[width=1\linewidth]{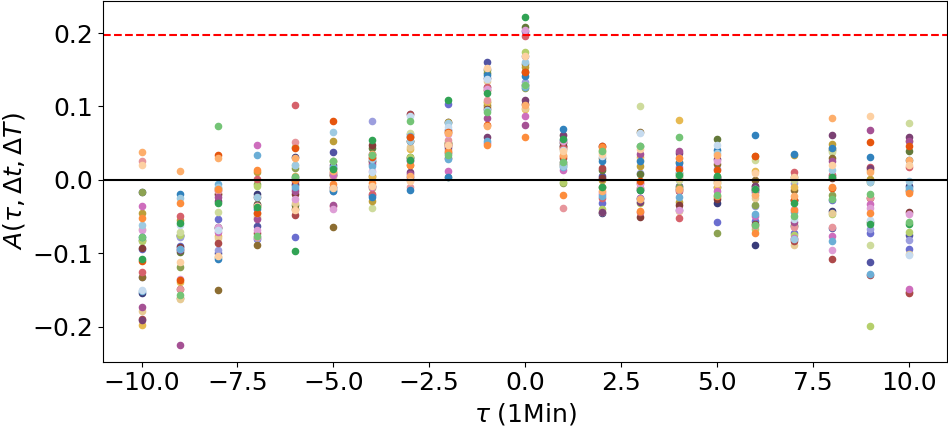}
    \caption{1-minute versus lagged 30-minute volatility.}
    \label{fig:asymmetry_1min_30min_norm}
\end{subfigure}%
\hspace{1em}
\begin{subfigure}[t]{.48\textwidth}
    \centering
    \includegraphics[width=1\linewidth]{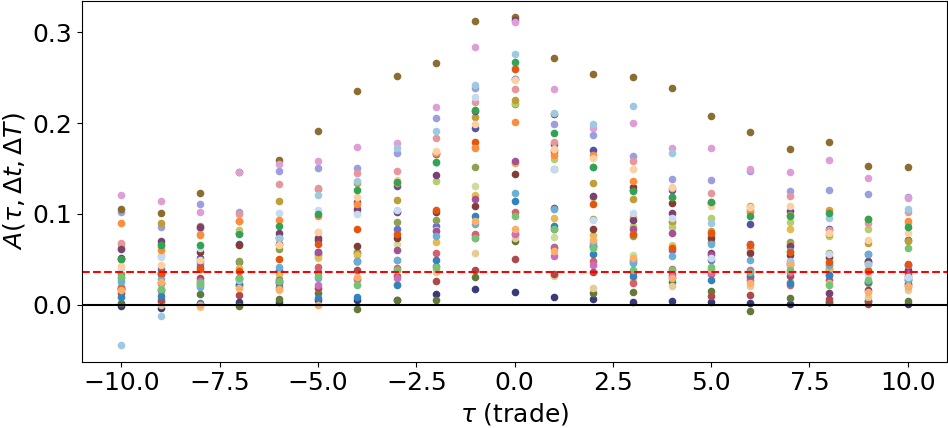}
    \caption{Trade versus lagged 1000-trade volatility.}
    \label{fig:asymmetry_trade_1000_norm}
\end{subfigure}%
\caption{Average fine-grained volatility versus lagged coarse-grained volatility --- Normalized.}
\label{fig:asymmetry_timescales_norm}
\end{figure}

Using the normalized returns $\hat{r}$, we conducted the same tests from our main paper as detailed in Section \ref{results}. The top-level result is that our main takeaways did not change; we found support for the same eight stylized facts with the normalized returns as the raw returns, and we did not find any additional support for the remaining three facts. Aside from the heavy tails analysis, which is discussed in Section \ref{heavy_tails}, the results on the normalized returns are shown in Figs. \ref{fig:autocorrelation_d_norm}, \ref{fig:skew_norm}, \ref{fig:abs_autocorrelation_norm}, \ref{fig:volume_volatility_norm}, \ref{fig:leverage_effect_norm}, and \ref{fig:asymmetry_timescales_norm}. For easy comparison, we re-display the main body (raw return) results here as well in Figs. \ref{fig:autocorrelation_raw}, \ref{fig:skew_raw}, \ref{fig:abs_autocorrelation_raw}, \ref{fig:volume_volatility_raw}, \ref{fig:leverage_effect_raw}, and \ref{fig:asymmetry_timescales_raw}. We see the magnitudes of some measures affected by the normalization process, such as the strength of correlation or skew. For the facts with a consistent signal across symbols, however, we see the same high-level signal here. For the facts without a clear signal (Facts \#3, 9, and 11), we similarly do not see a clear signal on the normalized returns.

\label{leverage_norm}

\label{volume_vol_norm}

\begin{figure}[H]
\centering
\begin{subfigure}[t]{.48\textwidth}
    \centering
    \includegraphics[width=1\linewidth]{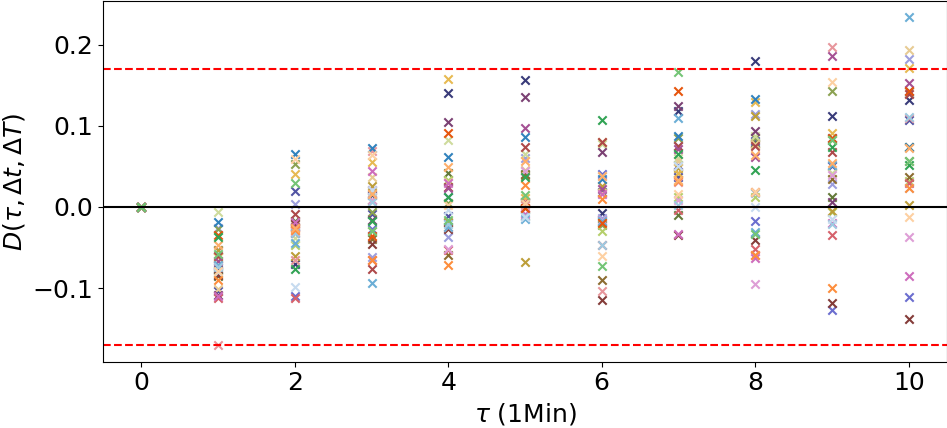}
    \caption{1-minute versus lagged 30-minute volatility.}
    \label{fig:asymmetry_1min_30min_diff_raw}
\end{subfigure}%
\hspace{1em}
\begin{subfigure}[t]{.48\textwidth}
    \centering
    \includegraphics[width=1\linewidth]{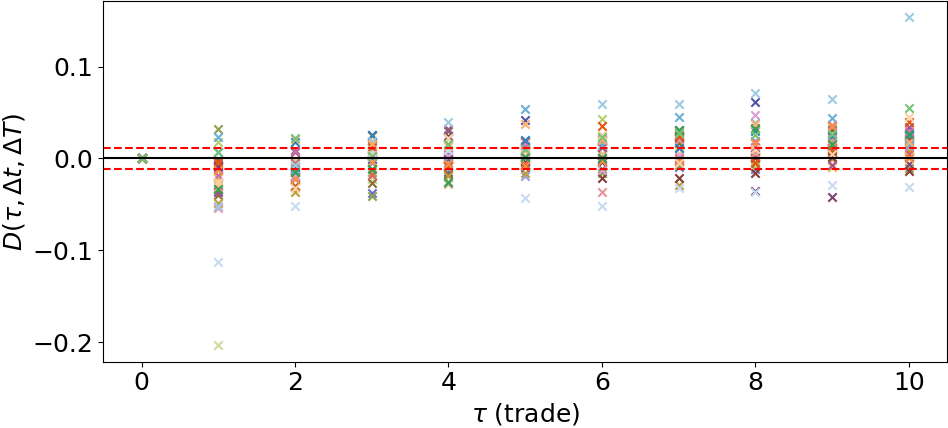}
    \caption{Trade versus lagged 1000-trade volatility.}
    \label{fig:asymmetry_trade_1000_diff_raw}
\end{subfigure}%
\caption{Asymmetry effect --- Difference between positive and negative lags for the correlation between the average fine volatility and lagged coarse volatility.}
\label{fig:asymmetry_timescales_diff_raw}
\end{figure}

\begin{figure}[H]
\centering
\begin{subfigure}[t]{.48\textwidth}
    \centering
    \includegraphics[width=1\linewidth]{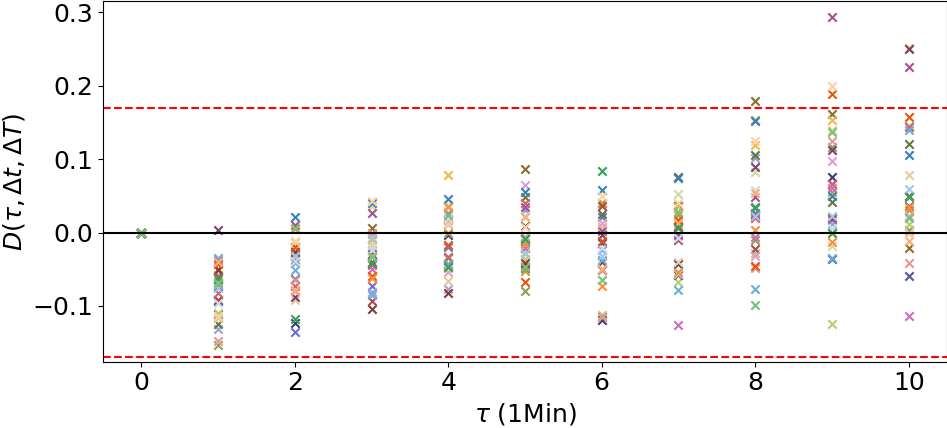}
    \caption{1-minute versus lagged 30-minute volatility.}
    \label{fig:asymmetry_1min_30min_diff_norm}
\end{subfigure}%
\hspace{1em}
\begin{subfigure}[t]{.48\textwidth}
    \centering
    \includegraphics[width=1\linewidth]{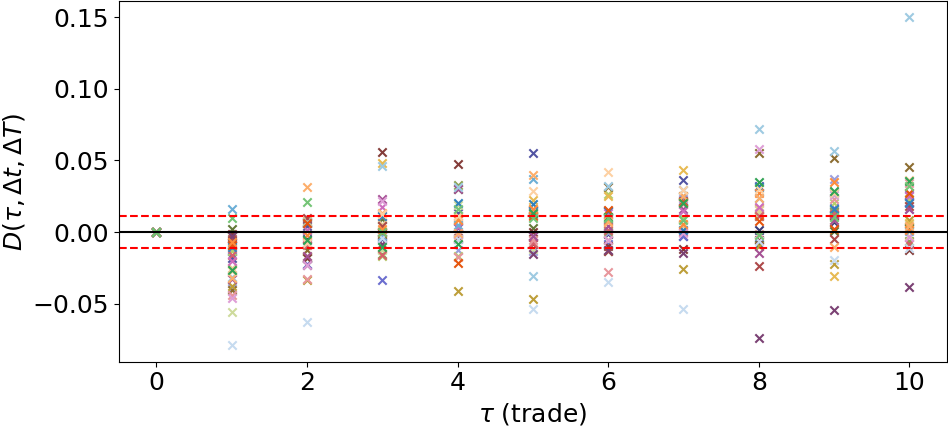}
    \caption{Trade versus lagged 1000-trade volatility.}
    \label{fig:asymmetry_trade_1000_diff_norm}
\end{subfigure}%
\caption{Asymmetry effect --- Difference between positive and negative lags for the correlation between the average fine volatility and lagged coarse volatility --- Normalized.}
\label{fig:asymmetry_timescales_diff_norm}
\end{figure}

\subsection{Autocorrelation of Trading Volume}
\label{volume_autocorr}

\begin{figure}[H]
\centering
\begin{subfigure}[t]{.48\textwidth}
    \centering
    \includegraphics[width=1\linewidth]{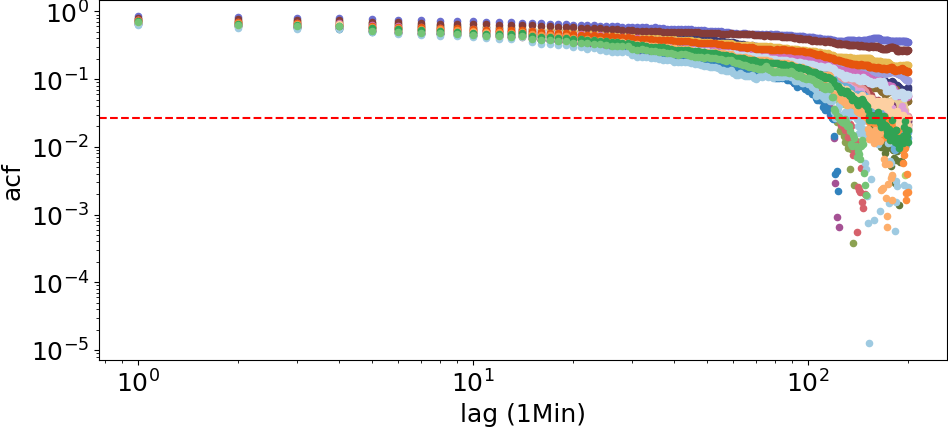}
    \caption{1-minute autocorrelation of trade volume.}
    \label{fig:trade_vol_autocorr}
\end{subfigure}%
\hspace{1em}
\begin{subfigure}[t]{.48\textwidth}
    \centering
    \includegraphics[width=1\linewidth]{1Min_sips_dow__D_Xvolatility_clustering_summary.png}
    \caption{1-minute autocorrelation of absolute (raw) returns.}
    \label{fig:vol_autocorr_1Min_comp}
\end{subfigure}%
\caption{Side-by-side comparison of the autocorrelation of trade volume and absolute returns at 1-minute timescale.}
\label{fig:trade_volume}
\end{figure}

In the main paper, we saw evidence of seven stylized facts exhibited in both the clock- and event-time, with some level of variation between these clocks depending on the specific measure. One possible factor that could influence the relationship between these two clocks (and any difference between them) is clustering of trading behavior in time. Similar to volatility clustering, we find evidence that the volume of trades in a minute clusters in time, and the autocorrelation of trading volume appears to exhibit a slow, power-law decay. This is shown in Fig. \ref{fig:trade_volume}. This is one avenue for future study for examining the relationship between clock-time and event-time in financial series.

\subsection{Supplementary Results}

\begin{figure}[H]
\centering
\begin{subfigure}[t]{.48\textwidth}
    \centering
    \includegraphics[width=1\linewidth]{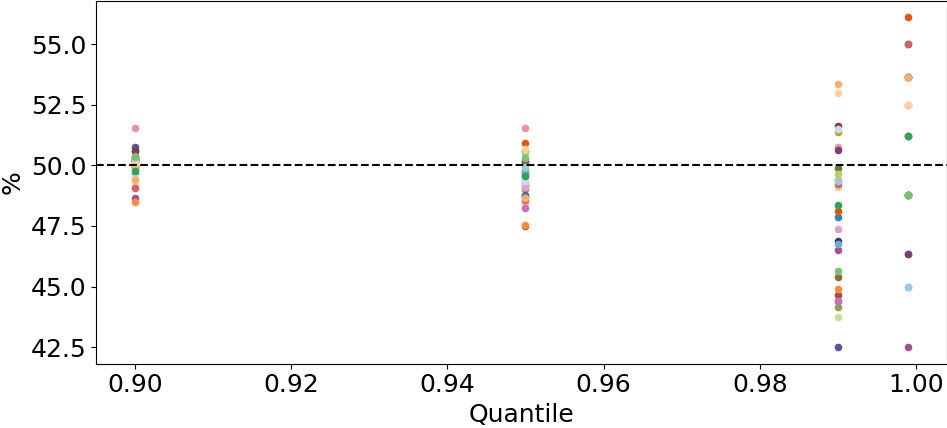}
    \caption{1-minute returns.}
    \label{fig:extreme_loss_perc_min}
\end{subfigure}%
\hspace{1em}
\begin{subfigure}[t]{.48\textwidth}
    \centering
    \includegraphics[width=1\linewidth]{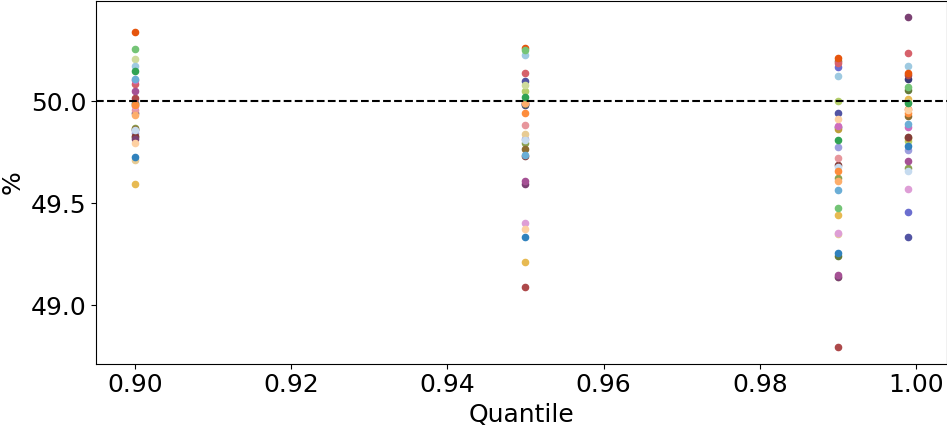}
    \caption{Trade-level returns.}
    \label{fig:extreme_loss_perc_trade}
\end{subfigure}%
\caption{Percentage of returns above a quantile-cutoff which were losses versus gains, in direct feeds perspective.}
\label{fig:extreme_loss_perc}
\end{figure}

In Fig. \ref{fig:extreme_loss_perc}, we show an alternative measure of gain/loss asymmetry (Fact \#3). Looking at the percentage of extreme returns that were gains versus losses, we did not see a disproportionate amount of large losses. This runs counter to the claims of Fact \#3, as noted in our main results (Section \ref{gain_loss}).

In Table \ref{table:ext_interarrival}, we give an alternative measure of intermittency, looking at the kurtosis of the interarrival times for extreme returns. Specifically looking at the 99$^{th}$-percentile largest returns, the interarrival times are heavy-tailed, with kurtosis values well above zero. This is an indication of intermittency, in addition to the results shown in Section \ref{intermittency}.
\begin{table}[]
\centering
\begin{tabular}{|l|r|r|}
\hline
Symbol &  1-min Return &  Trade Return \\
\hline
  AAPL &          7.14 &         85.66 \\
   AXP &         10.88 &        154.37 \\
    BA &          9.46 &        227.43 \\
   CAT &          9.56 &        110.16 \\
  CSCO &          7.98 &         40.35 \\
   CVX &          5.24 &        108.09 \\
   DIS &          8.81 &        118.30 \\
  DWDP &          3.88 &        153.06 \\
    GS &         11.24 &        129.76 \\
    HD &          6.50 &         96.56 \\
   IBM &          6.14 &        272.78 \\
  INTC &          5.81 &         38.07 \\
   JNJ &         13.65 &        126.76 \\
   JPM &          9.93 &        102.18 \\
    KO &          6.08 &         55.55 \\
   MCD &          9.11 &        153.49 \\
   MMM &          7.72 &        253.10 \\
   MRK &          6.33 &         86.95 \\
  MSFT &         15.36 &         75.55 \\
   NKE &          9.66 &        110.27 \\
   PFE &          6.58 &         47.73 \\
    PG &          7.35 &         60.79 \\
   TRV &          8.66 &        137.31 \\
   UNH &         11.11 &        184.69 \\
   UTX &          9.33 &        130.88 \\
     V &         10.44 &        117.30 \\
    VZ &          6.07 &         59.64 \\
   WBA &         10.11 &         62.48 \\
   WMT &          9.15 &        119.20 \\
   XOM &          6.21 &         66.06 \\
\hline
\end{tabular}
\caption{Kurtosis of the (intraday) interarrival times for 99$^{th}$-percentile extreme returns. This is an alternative measure of intermittency, as discussed in Section \ref{intermittency}.}
\label{table:ext_interarrival}
\end{table}

\begin{figure}[H]
\centering
\begin{subfigure}[t]{1\textwidth}
  \centering
  \includegraphics[width=1\linewidth]{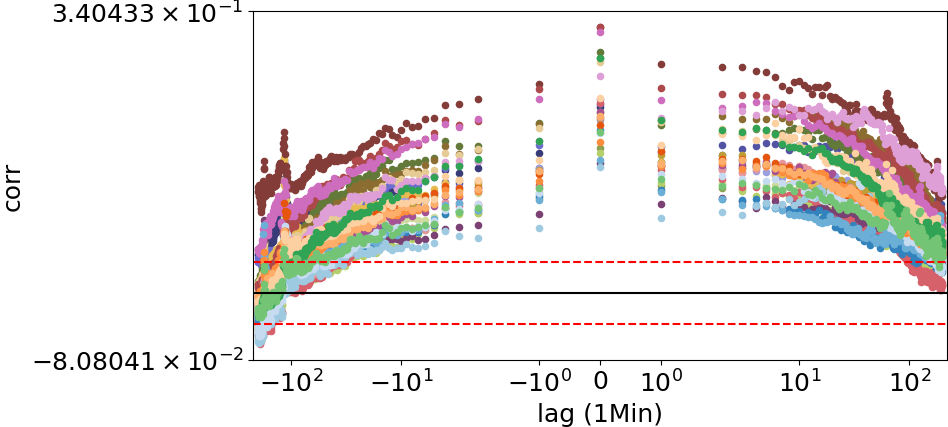}
\end{subfigure}
\caption{Volume of trades versus lagged volatility at $\Delta{t}=1Min$.}
\label{fig:volume_volatility_trades}
\end{figure}

In our main paper, we showed the relationship between share volume and volatility in both clock-time and event-time. In Fig. \ref{fig:volume_volatility_trades}, we show the relationship between trade volume and volatility, which looks roughly identical to share volume and volatility. Since share volume can vary in event-time but trade volume cannot, we used share volume as our volume metric in the main results in Section \ref{volume_vol}.

\subsection{Asymmetry in timescales - alternative measures}
\begin{figure}[H]
\centering
\begin{subfigure}[t]{0.48\textwidth}
    \centering
    \includegraphics[width=1\linewidth]{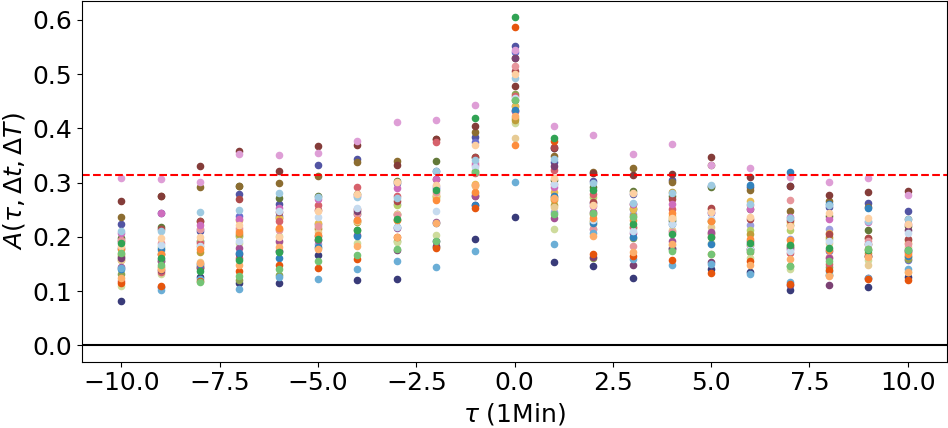}
    \caption{1-minute versus lagged 10-minute volatility.}
    \label{fig:asymm_10min_sub}
\end{subfigure}%
\hspace{1em}
\begin{subfigure}[t]{.48\textwidth}
    \centering
    \includegraphics[width=1\linewidth]{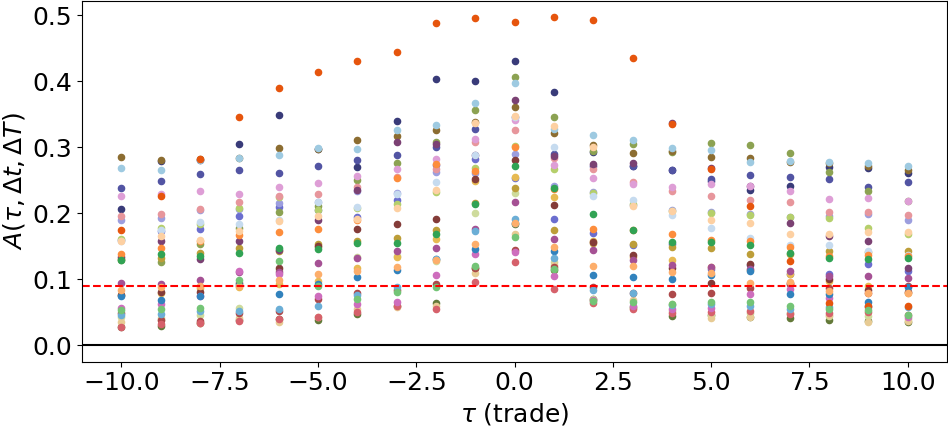}
    \caption{Trade versus lagged 100-trade volatility.}
    \label{fig:asymmetry_trade_100}
\end{subfigure}%
\caption{Asymmetry in timescales.}
\label{fig:asymm_supp}
\end{figure}

Fact \#11, asymmetry in timescales, expects coarse volatility measures to predict fine volatility better than the other way around. In our main results (Section \ref{asymmetry}), we used absolute returns at a coarse timescale $\Delta{T}$ and average absolute returns at a shorter timescale $\Delta{t}$ to test this effect. We specifically showed the results for $\Delta{T}=30Min$ and $\Delta{t}=1Min$ in clock-time and $\Delta{T}=1000\text{ trades}$ and $\Delta{t}=1\text{ trade}$ in even-time. In Fig. \ref{fig:asymm_supp}, we show alternative coarse timescales $\Delta{T}=10Min$ and $\Delta{T}=100\text{ trades}$, respectively.

\begin{figure}[H]
\centering
\begin{subfigure}[t]{0.48\textwidth}
    \centering
    \includegraphics[width=1\linewidth]{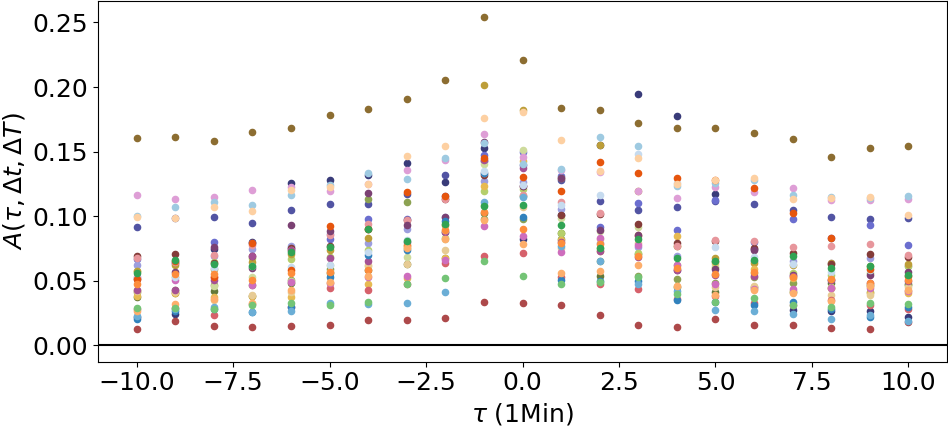}
    \caption{Rogers-Satchell volatility versus lagged 1-minute volatility.}
    \label{fig:rs_1}
\end{subfigure}%
\hspace{1em}
\begin{subfigure}[t]{.48\textwidth}
    \centering
    \includegraphics[width=1\linewidth]{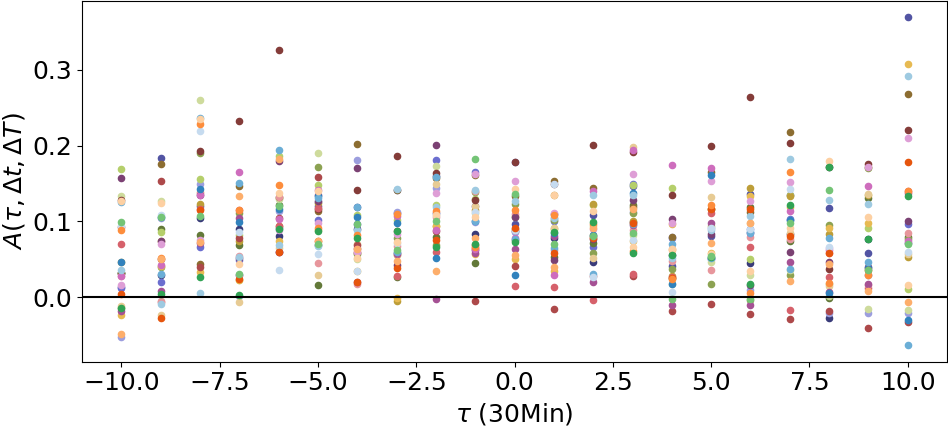}
    \caption{Rogers-Satchell volatility versus lagged 30-minute volatility.}
    \label{fig:rs_30}
\end{subfigure}%
\hspace{1em}
\begin{subfigure}[t]{0.48\textwidth}
    \centering
    \includegraphics[width=1\linewidth]{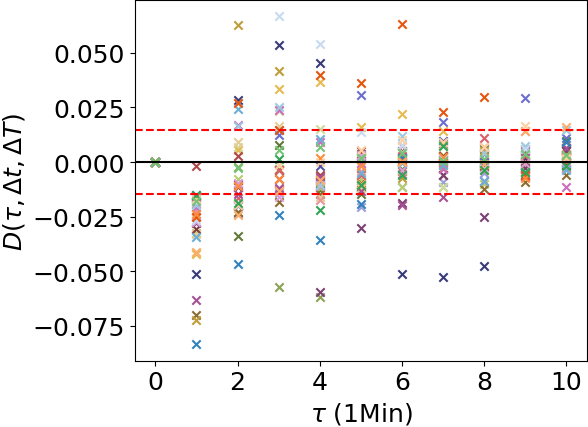}
    \caption{Asymmetry difference in correlation for positive and negative lags of Rogers-Satchell volatility versus lagged 1-minute volatility.}
    \label{fig:rs_diff_1}
\end{subfigure}%
\hspace{1em}
\begin{subfigure}[t]{.48\textwidth}
    \centering
    \includegraphics[width=1\linewidth]{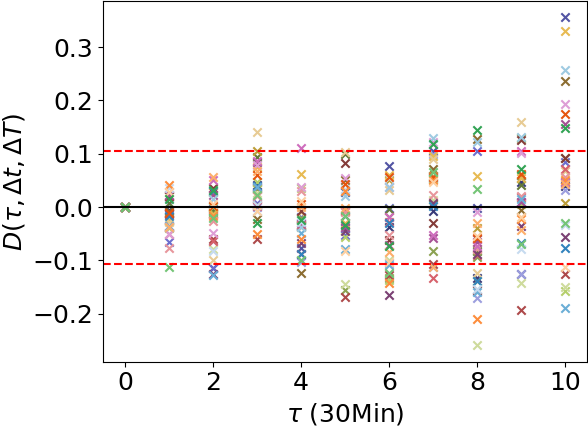}
    \caption{Asymmetry difference in correlation for positive and negative lags of Rogers-Satchell volatility versus lagged 30-minute volatility.}
    \label{fig:rs_diff_30}
\end{subfigure}%
\caption{Asymmetry in timescales using RS-volatility, in clock-time, raw returns.}
\label{fig:asymm_rs_clock_raw}
\end{figure}

\begin{figure}[H]
\centering
\begin{subfigure}[t]{0.48\textwidth}
    \centering
    \includegraphics[width=1\linewidth]{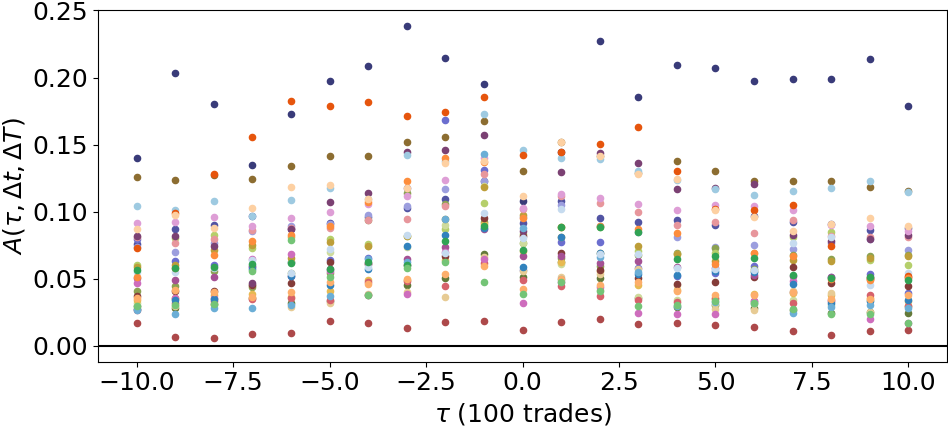}
    \caption{Rogers-Satchell volatility versus lagged 1-minute volatility.}
    \label{fig:rs_100}
\end{subfigure}%
\hspace{1em}
\begin{subfigure}[t]{.48\textwidth}
    \centering
    \includegraphics[width=1\linewidth]{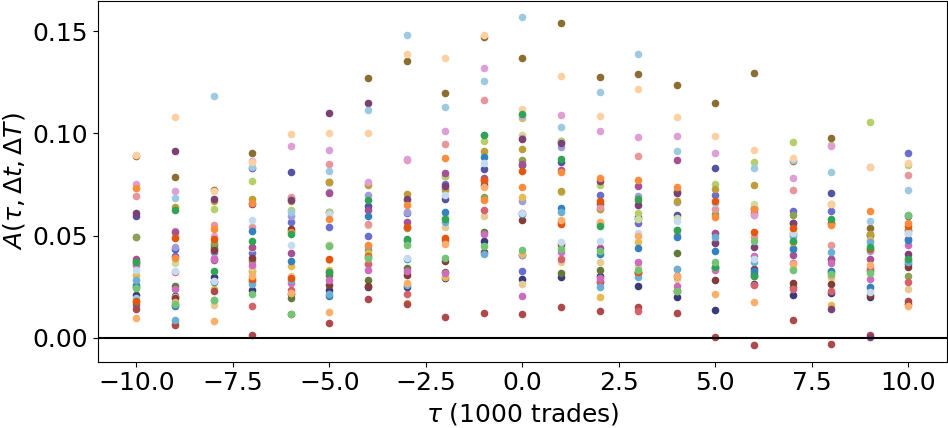}
    \caption{Rogers-Satchell volatility versus lagged 30-minute volatility.}
    \label{fig:rs_1000}
\end{subfigure}%
\hspace{1em}
\begin{subfigure}[t]{0.48\textwidth}
    \centering
    \includegraphics[width=1\linewidth]{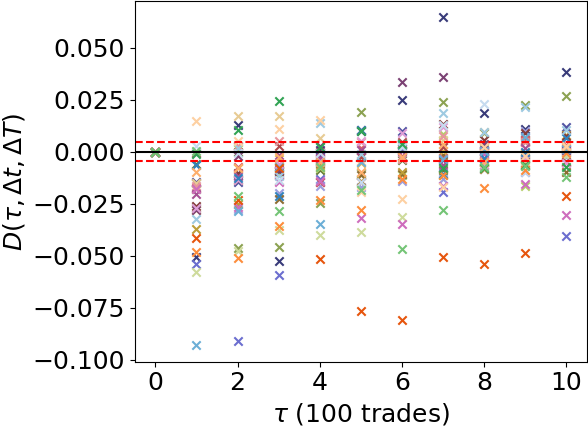}
    \caption{Asymmetry difference in correlation for positive and negative lags of Rogers-Satchell volatility versus lagged 1-minute volatility.}
    \label{fig:rs_diff_100}
\end{subfigure}%
\hspace{1em}
\begin{subfigure}[t]{.48\textwidth}
    \centering
    \includegraphics[width=1\linewidth]{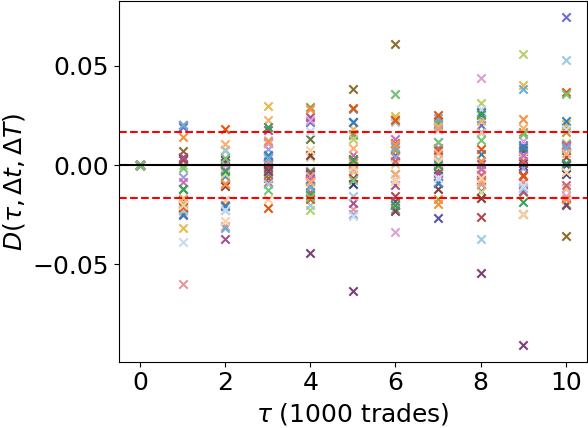}
    \caption{Asymmetry difference in correlation for positive and negative lags of Rogers-Satchell volatility versus lagged 30-minute volatility.}
    \label{fig:rs_diff_1000}
\end{subfigure}%
\caption{Asymmetry in timescales using RS-volatility, in clock-time, raw returns.}
\label{fig:asymm_rs_event_raw}
\end{figure}

We also considered an alternative fine volatility measure, using Rogers-Satchell volatility as done by Blanc et al. \cite{blanc_quadratic_2017}. For a time-bucket of size $\Delta{T}$, let $O_T$, $C_T$, $H_T$, and $L_T$ denote the open, close, high, and low prices, respectively. We consider the open price to be the price of the first trade in the bucket, the close to be the last, and any bucket that has no trades will forward fill the most recent price for each. The Rogers-Satchell volatility is then:
$$\sigma^{RS}\left(\Delta{T}\right)=\sqrt{\ln\left(H/O\right)\times\ln\left(H/C\right)+\ln\left(L/O\right)\times\ln\left(L/C\right)}$$
Our findings, shown in Fig. \ref{fig:asymm_rs_clock_raw} for clock-time and Fig. \ref{fig:asymm_rs_event_raw}, did not change the main takeaways for this fact discussed in Section \ref{asymmetry}. Similarly, we tested this on the normalized returns (using the normalization discussed in Section \ref{normailized}) and did not find this to alter our top-level results.

\begin{figure}[H]
\centering
\begin{subfigure}[t]{0.48\textwidth}
    \centering
    \includegraphics[width=1\linewidth]{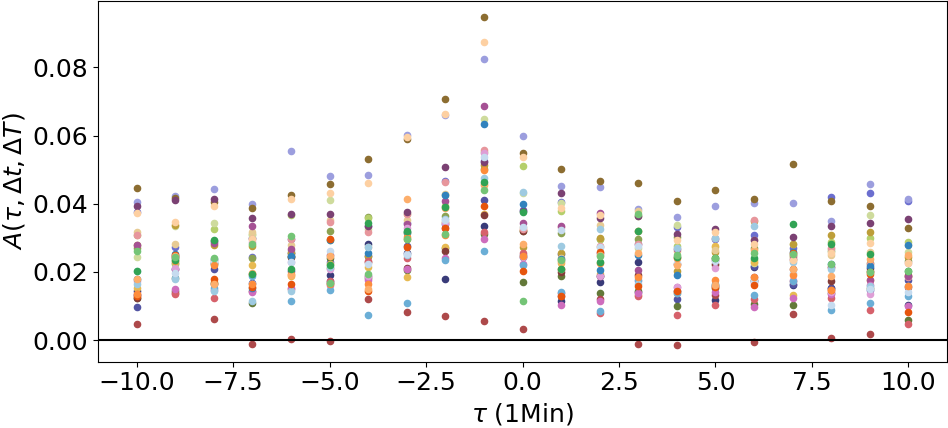}
    \caption{Rogers-Satchell volatility versus lagged 1-minute volatility --- normalized.}
    \label{fig:rs_1_n}
\end{subfigure}%
\hspace{1em}
\begin{subfigure}[t]{.48\textwidth}
    \centering
    \includegraphics[width=1\linewidth]{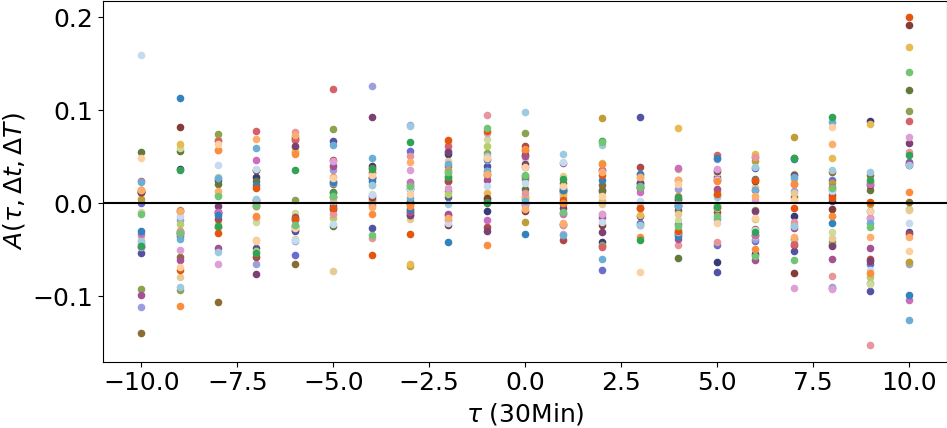}
    \caption{Rogers-Satchell volatility versus lagged 30-minute volatility --- normalized.}
    \label{fig:rs_30_n}
\end{subfigure}%
\hspace{1em}
\begin{subfigure}[t]{0.48\textwidth}
    \centering
    \includegraphics[width=1\linewidth]{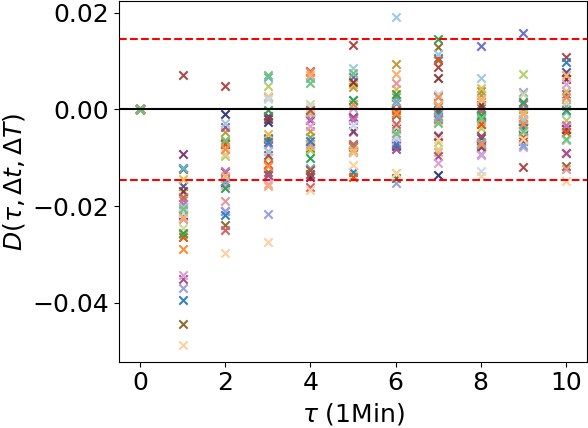}
    \caption{Asymmetry difference in correlation for positive and negative lags of Rogers-Satchell volatility versus lagged 1-minute volatility --- normalized.}
    \label{fig:rs_diff_1_n}
\end{subfigure}%
\hspace{1em}
\begin{subfigure}[t]{.48\textwidth}
    \centering
    \includegraphics[width=1\linewidth]{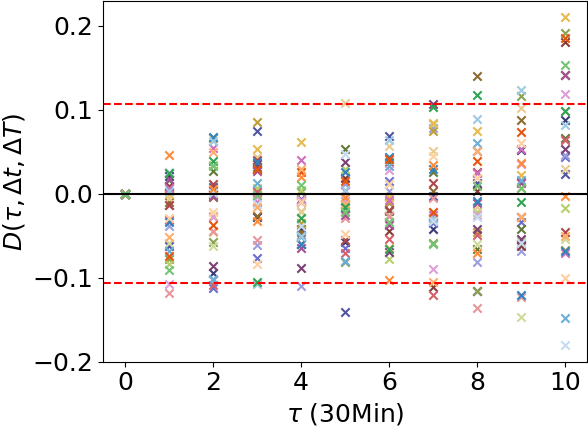}
    \caption{Asymmetry difference in correlation for positive and negative lags of Rogers-Satchell volatility versus lagged 30-minute volatility --- normalized.}
    \label{fig:rs_diff_30_n}
\end{subfigure}%
\caption{Asymmetry in timescales using RS-volatility, in clock-time, normalized returns.}
\label{fig:asymm_rs_clock_norm}
\end{figure}

\begin{figure}[H]
\centering
\begin{subfigure}[t]{0.48\textwidth}
    \centering
    \includegraphics[width=1\linewidth]{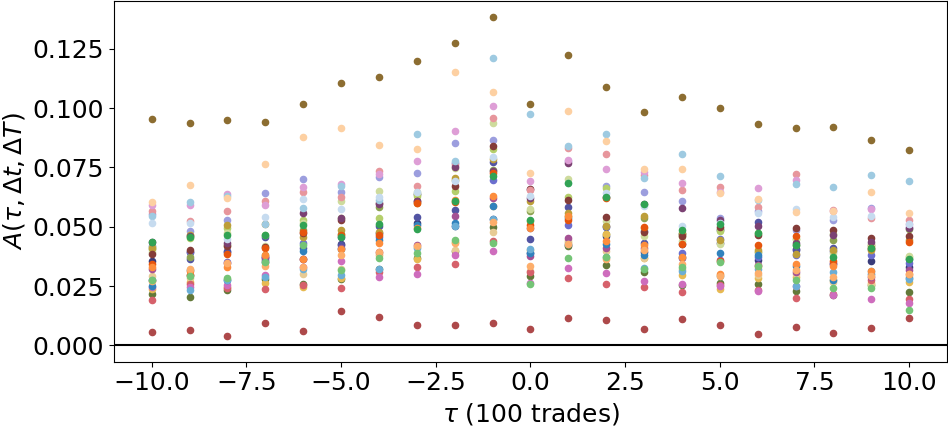}
    \caption{Rogers-Satchell volatility versus lagged 100-trade volatility --- normalized.}
    \label{fig:rs_100_norm}
\end{subfigure}%
\hspace{1em}
\begin{subfigure}[t]{.48\textwidth}
    \centering
    \includegraphics[width=1\linewidth]{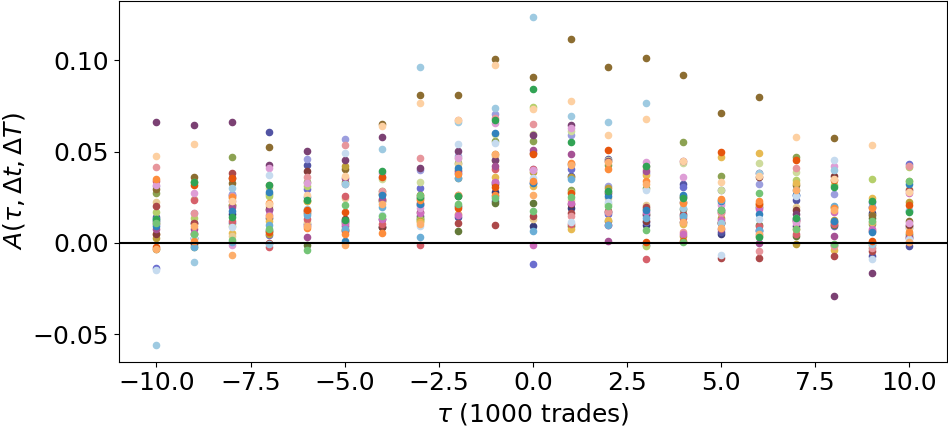}
    \caption{Rogers-Satchell volatility versus lagged 1000-trade volatility --- normalized.}
    \label{fig:rs_1000_norm}
\end{subfigure}%
\hspace{1em}
\begin{subfigure}[t]{0.48\textwidth}
    \centering
    \includegraphics[width=1\linewidth]{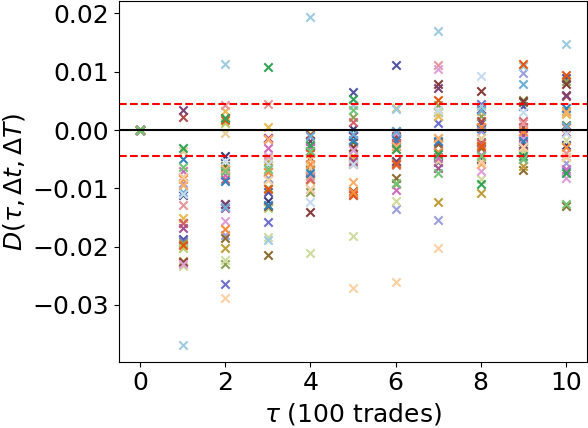}
    \caption{Asymmetry difference in correlation for positive and negative lags of Rogers-Satchell volatility versus lagged 100-trade volatility --- normalized.}
    \label{fig:rs_diff_100_norm}
\end{subfigure}%
\hspace{1em}
\begin{subfigure}[t]{.48\textwidth}
    \centering
    \includegraphics[width=1\linewidth]{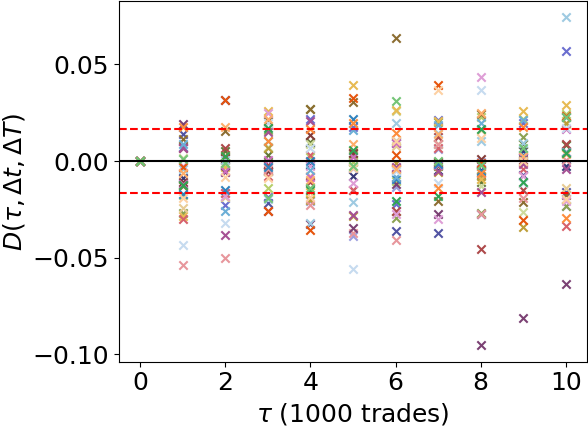}
    \caption{Asymmetry difference in correlation for positive and negative lags of Rogers-Satchell volatility versus lagged 1000-trade volatility --- normalized.}
    \label{fig:rs_diff_1000_norm}
\end{subfigure}%
\caption{Asymmetry in timescales using RS-volatility, in event-time, normalized returns.}
\label{fig:asymm_rs_event_norm}
\end{figure}

\subsection{Interarrival times}
\label{interarrival}
\begin{table}
\centering
\begin{tabular}{|l|r|r|r|r|r|}
\hline
symbol & mean & median & std dev & min & max \\
\hline
AAPL & 92,560.25 & 542.00 & 290,519.52 & 0 & 132,293,995 \\
AXP & 696,067.68 & 822.00 & 2,580,901.55 & 0 & 1,934,000,552 \\
BA & 375,495.40 & 1,143.00 & 1,372,434.04 & 0 & 1,069,737,369 \\
CAT & 462,875.38 & 648.00 & 2,069,858.46 & 0 & 1,705,419,450 \\
CSCO & 194,259.59 & 147.00 & 1,008,009.91 & 0 & 1,487,856,230 \\
CVX & 448,976.42 & 1,002.00 & 1,595,933.35 & 0 & 1,051,137,673 \\
DIS & 372,489.70 & 1,113.00 & 1,181,464.65 & 0 & 649,561,007 \\
DWDP & 327,602.60 & 336.00 & 1,471,429.09 & 0 & 1,531,255,491 \\
GS & 583,218.27 & 765.00 & 2,288,273.76 & 0 & 1,538,800,294 \\
HD & 454,093.38 & 1,022.00 & 1,748,204.14 & 0 & 1,111,566,254 \\
IBM & 486,224.82 & 1,223.00 & 1,760,146.58 & 0 & 1,020,254,692 \\
INTC & 168,821.24 & 141.00 & 893,994.30 & 0 & 1,659,096,150 \\
JNJ & 356,560.55 & 1,049.00 & 1,367,533.53 & 0 & 977,249,817 \\
JPM & 210,493.29 & 454.00 & 942,723.39 & 0 & 1,199,807,075 \\
KO & 313,075.04 & 283.00 & 1,206,901.39 & 0 & 1,295,475,229 \\
MCD & 642,079.27 & 1,473.00 & 2,132,398.26 & 0 & 999,165,025 \\
MMM & 857,397.83 & 1,084.00 & 3,620,091.00 & 0 & 2,766,754,265 \\
MRK & 315,085.83 & 435.00 & 1,382,843.25 & 0 & 1,252,446,928 \\
MSFT & 99,382.11 & 207.00 & 474,386.70 & 0 & 1,150,040,313 \\
NKE & 484,302.67 & 536.00 & 1,993,847.96 & 0 & 2,280,079,800 \\
PFE & 222,415.54 & 236.00 & 1,321,824.69 & 0 & 2,823,418,004 \\
PG & 337,284.41 & 445.00 & 1,522,517.88 & 0 & 1,272,005,190 \\
TRV & 1,370,633.87 & 1,131.00 & 4,477,760.52 & 0 & 2,363,298,851 \\
UNH & 546,788.73 & 1,030.00 & 2,006,331.16 & 0 & 1,310,814,029 \\
UTX & 546,987.42 & 999.00 & 2,587,718.60 & 0 & 1,903,482,893 \\
V & 306,346.81 & 830.00 & 1,359,195.94 & 0 & 1,924,749,052 \\
VZ & 269,117.87 & 334.00 & 1,303,559.17 & 0 & 2,147,190,495 \\
WBA & 507,570.75 & 336.00 & 2,335,817.45 & 0 & 2,563,291,850 \\
WMT & 373,781.26 & 651.00 & 1,521,786.38 & 0 & 2,125,795,062 \\
XOM & 280,564.92 & 396.00 & 1,052,449.64 & 0 & 557,360,208 \\
\hline
all & 423,418.43 & 693.77 & 1,695,695.21 & 0.00 & 1,526,780,108.10 \\
\hline
\end{tabular}
\caption{Intraday interarrival times of trades for Dow 30 stocks in our sample.}
\label{table:interarrival_stats}
\end{table}

\end{document}